# Ho$^{3+}$-doped CALGO crystals for high-power ultrafast 2.1-μm lasers

ANNA SUZUKI,[1] PAVEL LOIKO,[2] WEICHAO YAO,[1] PARISA BAGHERY,[1] MARTIN HOFFMANN,[1] KIRILL EREMEEV,[2] PATRICE CAMY,[2] ALAIN BRAUD,[2] SERGEI TOMILOV,[1] YICHENG WANG,[1] AND CLARA J. SARACENO[1,3*]

[1]*Photonics and Ultrafast Laser Science, Ruhr-Universität Bochum, Universitätsstrasse 150, 44801 Bochum,Germany*
[2]*Centre de Recherche sur les Ions, les Matériaux et la Photonique (CIMAP), UMR 6252 CEA-CNRS-ENSICAEN, Université de Caen Normandie, 6 Boulevard Maréchal Juin, 14050 Caen Cedex 4, France*
[3]*Research Center Chemical Science and Sustainability, University Alliance Ruhr, 44801 Bochum, Germany*
*\*clara.saraceno@ruhr-uni-bochum.de*

**Abstract:** Ho$^{3+}$-doped disordered CaAlGdO$_4$ (CALGO) crystals have recently emerged as a promising gain material platform for next-generation high-power ultrafast 2.1-μm laser systems. This laser gain material offers a unique combination of high-gain, small quantum defect, inhomogeneously broadened spectra, and good thermal conductivity, enabling ultrashort pulse generation and amplification at high-average power and high pulse energy. Many systems, including mode-locked oscillators and amplifiers with state-of-the-art performance, have been demonstrated in the last few years that promise to meet growing application demands for efficient ultrafast laser technology in this wavelength region. In this review paper, we summarize recent achievements using this gain material both in oscillators and amplifiers and place these results in the state-of-the-art of 2-μm ultrafast laser technology, present detailed spectroscopic characterization of this material, and discuss future perspectives of further performance scaling of Ho:CALGO lasers.

## 1. Introduction

High-power ultrafast lasers in the short-wavelength infrared (SWIR) range (typically defined as the 1.7-2.5 μm spectral region) offer promising perspectives in a wide range of scientific and industrial applications. For example, they enable nonlinear processing of narrow bandgap semiconductor materials such as silicon and germanium, which are opaque in the near-infrared range [1,2], opening the door to in-volume or through-substrate material modification. Moreover, they are in high demand as driving sources for nonlinear frequency conversion, for example, to access the extreme ultraviolet (XUV) and soft X-ray regions via high-harmonic generation [3], which will be a key tool for several applications, for example, in time-resolved studies of electron and nuclear dynamics [4]. SWIR sources also serve as efficient drivers for down-conversion to the mid-infrared (MIR) region, using non-oxide nonlinear crystals [5]. The MIR region includes the molecular fingerprint region, which is important for molecular spectroscopy, chemical sensing, and environmental monitoring [6,7]. Furthermore, SWIR sources enable efficient terahertz generation based on optical rectification [8] or two-color plasma schemes [9], enabling access to a spectral range useful for non-destructive imaging [10], and spectroscopy [11–13].

Traditionally, this spectral region was only accessible at high energy and high power using parametric conversion of more powerful near-infrared drivers, i.e. using parametric oscillators and amplifiers. For example, in the SWIR spectral range, a record-high average power of 52 W was achieved at a 2.1-μm wavelength and 52.6-kHz repetition rate via 5-stage OPAs driven by a 500-W Yb-based disk amplifier system [14]. The highest pulse energy over 50 mJ was reported at 2.44-μm and 10-Hz repetition rate, using a dual-chirped OPA technique with a high-energy Ti:sapphire laser driver [15]. In the kilohertz repetition rate range, a pulse energy of 4.9 mJ was obtained at 2.1-μm via Yb-based disk amplifier driven 3-stage optical parametric chirped pulse amplifier at 10 kHz [16]. However, such laser systems are complex and expensive, and remain nearly exclusively dedicated to specialized scientific use, in institutions where advanced expertise in ultrafast laser systems operation is present. This has prevented the widespread deployment of laser systems in this spectral region, and therefore, slowed down many potential areas of application.

In contrast, efficient 2-µm laser sources based on direct emission and amplification using different active ions and host materials have also seen significant progress in the last decades, offering a promising alternative to reach this spectral region with higher efficiency and generally more compact, economical, and simple systems. Various mode-locked laser oscillators and ultrafast amplifiers have been demonstrated in the last decades using $Tm^{3+}$-, $Ho^{3+}$-, and $Cr^{2+}$-doped gain materials, which we detail in the state-of-the-art section below. Progress in these laser systems has been fueled by advances in gain materials [17], power scaling and commercialization of pump sources [18,19], and more mature coating technologies for this wavelength range. Figure 1 shows an overview of high-power ultrafast laser sources in the SWIR region defined above, including record-holding OPAs and OPOs for comparison. From this overview graph, it becomes apparent that this is an active area of research with large potential for further developments.

In this paper, we review the recent achievements of one particularly promising material platform that has emerged as promising for average power scaling and ultrafast laser operation at 2.1 µm: ultrafast Ho:CALGO laser systems. Firstly, we present a detailed state-of-the-art of lasers emitting in this spectral region, with emphasis on ultrafast lasers with average power scaling potential, and place recent progress of Ho:CALGO lasers in this context. Then we present measurements of the physical and spectroscopic properties of Ho:CALGO. Subsequently, we extensively summarize the most recent state-of-the-art demonstrations of mode-locked laser oscillators and chirped pulse amplifiers (CPAs) based on Ho:CALGO and present new data on prospects for reaching shorter pulse durations towards few-cycles. Finally, we discuss future prospects for Ho:CALGO-based laser technology towards next-generation 2 µm high-power ultrafast laser sources.

## 2. State of the art of high-power lasers at 2.1 µm

In this section, we summarize the most recent demonstrations of ultrafast laser systems emitting in the SWIR spectral region and present the main advantages and disadvantages of competing technologies. We note that these advantages and disadvantages are relative to the targeted application field. We focus on systems aiming to achieve high average power and short pulse durations. Additionally, note that other gain media are also emerging with strong potential, such as semiconductor-gain based mode-locked vertical external-cavity surface-emitting lasers (VECSELs) [20], that could potentially also become competitive in the near future. We leave this technology out for the purpose of this review and focus on ion-doped crystalline materials or glasses. For clarity, we divide the text by active-ion used for laser emission and for better illustration of the advantages of each material family, and to compare with Ho:CALGO, we present in Fig. 2 and Table 1 an overview of relevant laser material properties for some materials that have seen particularly pronounced progress in the last years.

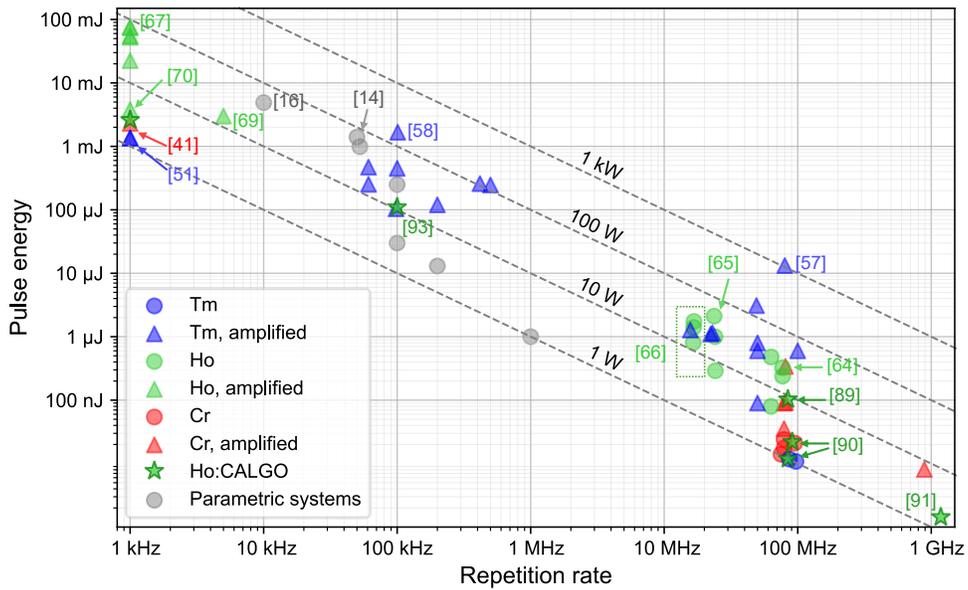

Fig. 1. Overview of high-power ultrafast laser oscillators and amplifiers in the SWIR wavelength region, and optical parametric conversion systems (The dataset includes systems with average output powers >1 W and pulse durations <5 ps.).

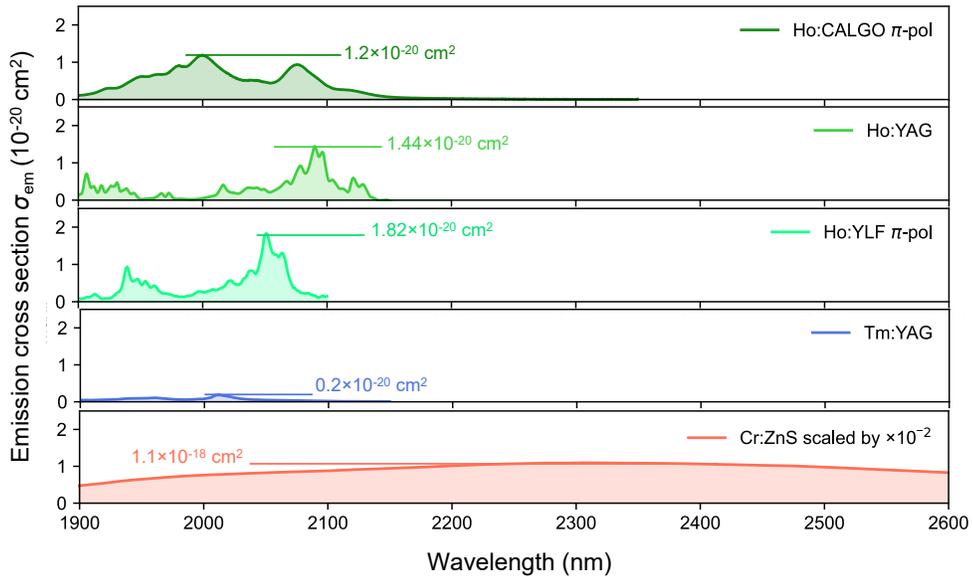

Fig. 2. Comparison of emission cross section spectra of Ho:CALGO, Ho:YAG [21], Ho:YLF [22], Tm:YAG [23], and Cr:ZnS [24].

Table 1. Optical and thermal properties of representative gain media for SWIR lasers*

|  | Cr:ZnS/ZnSe | Tm:YAG | Tm:YLF | Tm:silica | Ho:YAG | Ho:YLF | Ho:CALGO |
|---|---|---|---|---|---|---|---|
| Upper laser level | 0.005/0.0054 [25] | 9.42 (8 at.%) | 15.25 (0.5 at.%) | 0.515 (0.3 mol.%) | 7** [28] | 16 (0.5 at.%) | 5.27 (1% doped) |

| | | | | | | |
|---|---|---|---|---|---|---|
| lifetime (ms) | | doped) [23] | doped) [26] | doped) [27] | doped) [29] | |
| Thermal conductivity (Wm$^{-1}$K$^{-1}$) | 27 [30]/23.2 [31] | 12.9 [32] | 5.3∥$a$ 7.2∥$c$ [33] | 1.3-1.4 [34] | - | 9.5 [35] |
| Thermo-optic coefficient dn/dT (10$^{-6}$ K$^{-1}$) | 49.8/59.7 [36] | 9 [37] | -4.6∥$a$, -6.6∥$c$ [33] | 8.16-8.53 [38] | - | -7.6∥$a$ -8.6∥$c$ [39] |

*All values are given at room temperature (except for the lifetime of Ho:YAG measured at 20 K**), and the reported thermo-optic coefficients and thermal conductivities are for undoped materials.

*Cr-doped materials*

Cr$^{2+}$-doped chalcogenide materials emitting around 2.3-2.4 μm, often referred to as the 'mid-infrared Ti:sapphire', have the large advantage of an ultra-broad gain bandwidth and large gain cross sections due to vibronic transitions, as shown in Fig. 2. These materials are excited in the 1.5-1.9 μm range, typically using Er- or Tm-based lasers or laser diodes. Due to their exceptionally broad gain bandwidth, it is possible to directly generate few-cycle pulses with hundreds of milliwatt to few-watt level output powers from mode-locked lasers [24,25]. However, Cr$^{2+}$-doped materials suffer from very short upper-level lifetimes on the order of a few microseconds, which severely limits their energy storage capability. Therefore, they are poorly suited for CW-pumped laser amplifiers at kilohertz repetition rates and require pulsed pumping. In addition, their relatively large quantum defects (20-40%) impose severe thermal constraints for power scaling, requiring complex cooling strategies for reaching 100-W power levels [40]. High-energy laser operation in the CPA arrangement has been demonstrated at rather low repetition rates, e.g. 1-kHz Cr:ZnSe multi-pass amplifier generated 2.3-mJ pulse energy with 88-fs pulse duration [41], and close to terawatt level peak power was achieved with 127.6-fs pulse duration at a lower repetition rate of 1 Hz [42]. However, the high nonlinear refractive index $n_2$ [43,44] and high thermo-optic coefficients *dn/dT* of ZnS/ZnSe present an additional challenge for further energy scaling, which requires suppressing unwanted nonlinearity and avoiding crystal damage.

*Tm-doped materials*

Among all the different technologies that have advanced laser systems directly emitting in the SWIR spectral region, Tm$^{3+}$-doped laser materials have shown fast-paced progress. Laser systems based on different architectures (fiber, slabs and bulk) and in different hosts (glasses, crystals, ceramics) have been demonstrated [17,45,46]. One of the most attractive aspects of Tm is the possibility of diode-pumping using widely available, high-power 0.79-μm diodes. This pumping scheme benefits from the so-called two-for-one process via cross-relaxation in which one pump photon can be recycled to yield two emitted laser photons, which allows to increase quantum efficiency up to 2 and to obtain efficient laser operation, overcoming the low Stokes efficiency of less than 40% [47]. Alternatively, 1.6-μm in-band pumping with Er-doped fiber lasers, Raman-shifted Er fiber lasers, or laser diodes is possible, which benefits from a smaller quantum defect and the unnecessary use of high doping concentrations [48,49], but comes with a higher cost and lower available power levels.

Among the different hosts for the Tm$^{3+}$ ion, crystalline materials exhibit relatively broad gain bandwidths, which allow us to directly generate tens to hundreds of femtosecond pulses from mode-locked lasers. Ultrashort pulse generation has been extensively demonstrated in a variety of Tm$^{3+}$-doped bulk materials using different mode-locking techniques, with average power levels ranging from tens to hundreds of mW [50]. Moreover, high-energy ultrafast laser amplifier systems have also been demonstrated. A regenerative amplifier (RA) based on bulk

Tm:YAP achieved 1.35 mJ of amplified pulse energy with 265-fs pulse duration at a 1-kHz repetition rate [51]. Despite these achievements, Tm-based laser systems generally suffer from significant thermal load due to a relatively large quantum defect, which represents a primary challenge for power scaling. In fact, in the Tm:YAP RA [51], the gain crystal was actively cooled to −19.3°C using a thermoelectric cooling unit to mitigate thermal effects, indicating the need for advanced cooling strategies when scaling toward higher average powers with bulk geometry.

$Tm^{3+}$-doped fiber lasers, on the other hand, benefit from the high transparency of silica glass at Tm emission wavelengths and from the intrinsic advantages of the superior cooling efficiency of the fiber geometry. A wide variety of mode-locked Tm-doped fiber oscillators have been demonstrated using different mode-locking techniques [52,53]. Due to the small mode area and the relatively low pulse energy that can be sustained in the fiber, their average output power is typically low—ranging from a few milliwatts to tens of milliwatts—compared with bulk crystalline lasers. Nevertheless, the broad gain bandwidth of Tm-doped glass has enabled the generation of very short pulse durations [54–56], establishing these oscillators as attractive seed sources for ultrafast systems. Furthermore, the excellent cooling efficiency of the fiber geometry has enabled remarkable progress in high-average-power ultrafast laser systems using chirped-pulse amplification (CPA) in recent years. For example, 1-kW average power was demonstrated for the first time with a Tm-fiber amplifier at an 80-MHz repetition rate [57], and 1.3-mJ pulse energy was achieved at 100 kHz with the coherent beam combining technique [58] to overcome its high nonlinearity due to the small mode area, illustrating the challenges for energy scaling.

It is also worth mentioning that the operation wavelength of Tm-based laser systems strongly depends on the inversion level. Tm lasers operating at relatively low inversion levels—such as most oscillators—typically emit at wavelengths above 2 µm and are therefore not significantly affected by water-vapor absorption in air. In contrast, Tm lasers operated at high inversion levels, as commonly required in amplifiers, typically emit around 1.9 µm, where strong water-vapor absorption leads to losses and beam degradation during free-space propagation. To mitigate these effects, research on the development of Tm-based amplifiers at longer wavelengths is increasing, where atmospheric absorption is substantially reduced [59–61]. However, enforcing amplification at longer wavelengths can only use lower gain and poses challenges for efficient energy extraction.

*Tm,Ho-doped materials*

$Tm^{3+}$,$Ho^{3+}$-codoped materials have attracted attention as a potential solution for generating shorter pulses. In these materials, the emission bands of $Tm^{3+}$ and $Ho^{3+}$ partially overlap, producing a broadened effective gain spectrum that can support shorter pulse durations compared to $Tm^{3+}$ or $Ho^{3+}$ singly doped lasers. They can be pumped via the $Tm^{3+}$ absorption band, and the $Ho^{3+}$ ions are indirectly excited by energy transfer from $Tm^{3+}$ ions. This pump scheme allows the use of diode lasers or Ti:sapphire laser pumping. Remarkably, mode-locked Tm,Ho-codoped lasers achieved pulse durations as short as 26 fs [62], which is comparable to mode-locked $Cr^{2+}$-doped lasers. However, the efficiency of Tm,Ho-codoped systems is limited by imperfect energy transfer from Tm to Ho and additional energy upconversion channels between $Tm^{3+}$ and $Ho^{3+}$ ions. These processes reduce the overall laser efficiency [17,63], making power scaling challenging, and consequently, more than watt-level laser sources have not been reported so far.

*Ho-doped materials*

In this context, $Ho^{3+}$-doped materials provide an attractive solution to several of the fundamental limitations for power and energy scaling encountered in $Tm^{3+}$- and $Cr^{2+}$-doped materials. $Ho^{3+}$-doped materials offer relatively large gain cross sections and long upper-level lifetimes on the millisecond level, which ensure large energy storage capability. These materials are typically in-band pumped at 1.9-µm, resulting in a quantum defect of less than 10%, which is significantly lower than that of $Tm^{3+}$- and $Cr^{2+}$-doped materials. Moreover, $Ho^{3+}$-doped lasers naturally emit around 2.1 µm, which lies in an atmospheric transmission window, and are suited for free-space propagation of high-power laser beams. Compared to $Tm^{3+}$-doped materials, however, the maximum doping concentration is generally much lower to avoid quenching due to energy transfer upconversion (ETU), which will be discussed in detail in the following sections. Consequently, Ho laser systems typically use lower doping concentrations with longer crystal lengths to achieve sufficient pump absorption. In this regard, the use of in-band Tm-fiber laser pumping is attractive since it offers excellent beam quality at high power, therefore, pump and laser can propagate almost collinearly, allowing the use of long crystal lengths without sacrificing mode-matching, making them highly promising for high-power, high-energy laser operation. The main challenge in Ho-doped laser systems is the typically narrow and structured gain cross-sections, making it challenging to reach very short pulse durations in both oscillators and amplifiers. Nevertheless, ultrafast laser development has seen remarkable progress in recent years. For example, mode-locked lasers based on Ho:YAG thin disks showed excellent performance. The first ultrafast Ho:YAG thin-disk laser was demonstrated in [64], achieving 25-W average output power with a 270-fs pulse duration using Kerr-lens mode-locking (KLM). Subsequently, the highest average power of any mode-locked laser in this spectral band with 50 W was achieved in a Ho:YAG thin disk laser oscillator using a semiconductor saturable absorber mirror (SESAM) with a pulse duration of 1.13 ps [65], and KLM enabled shorter pulse durations of 434 fs and 350 fs with average powers of 29 W and 23 W, respectively [66]. In addition, high-energy ultrafast laser amplifiers have also been demonstrated, mainly based on Ho:YLF. A pulse energy of 75 mJ with a pulse duration of 2.2 ps was demonstrated by a Ho:YLF RA followed by two booster stages at 1 kHz [67], and a cryogenically cooled Ho:YLF amplifier achieved 260-mJ pulse energy with a 16-ps pulse duration [68]. This illustrates the enormous potential of Ho-doped laser systems for high-power and high-energy operation, but also shows the bottleneck of their narrow and highly structured gain profiles as mentioned above. For example, the shortest pulse duration of conventional Ho-doped lasers was 220 fs which was generated from KLM Ho:YAG thin-disk lasers [64]. In contrast to Tm- and Cr-lasers, which benefit from broader and smoother gain profiles, sub-100-fs generation is typically challenging for mode-locked Ho lasers, and sub-ps pulse amplification is almost impossible unless a spectral and/or phase shaping technique is utilized [69,70].

Disordered host materials offer an attractive solution to this matter. The disordered crystal structure causes inhomogeneous spectral broadening on the transition lines of active ions. This makes the spectral profile broad and flat, which is highly favorable for ultrashort pulse generation and amplification. This however, often comes with a significant decrease in thermal conductivity. In this regard, $CaAlGdO_4$ (CALGO) and its family of disordered calcium aluminates exhibit exceptionally high thermal conductivity [39,71]. Therefore, the unique combination of a broad and flat gain profile and good thermo-mechanical properties makes CALGO a particularly promising platform for high-power ultrafast laser development. Early on, Yb-doped CALGO has been utilized for ultrafast 1-µm laser systems in bulk and thin-disk geometries [72]. Remarkably, Kerr-lens mode-locked Yb:CALGO oscillators enabled few-cycle pulse generation [73,74], whose pulse durations are comparable to Ti:sapphire lasers, which have been the gold standard for ultrafast laser systems. In addition, CPA using Yb-doped disordered aluminates enabled millijoule amplification with pulse durations in the 100-fs to sub-100-fs regime [75–79], which is suitable for high-peak-power applications. CALGO also worked well with other active dopants, such as $Tm^{3+}$ or $Tm^{3+}$, $Ho^{3+}$-codoping, especially for broadband mode-locked lasers [80–87]. However, their output power has so far remained rather low, typically limited to below the watt level. In contrast, $Ho^{3+}$-doped CALGO is promising for power scaling. It offers the combination of the favorable properties of Ho$^{3+}$ laser transitions, an efficient in-band pumping scheme, and a broad and flat spectral profile created by the disordered host crystal structure [88], making it a convincing platform for high-power ultrafast laser development. In the last few years, record high-power bulk lasers have been demonstrated with this material (up to 8.7 W and 369-fs pulses using SESAM mode-locking [89], up to 2 W with <100 fs pulse duration with KLM [90–92]) and regenerative amplification with sub-ps pulses with high pulse energies and high repetition rate [93]. We detail these results in more depth in Section 4.

## 3. Optical spectroscopy of Ho:CALGO

The excellent laser performance demonstrated with Ho:CALGO highlights its strong potential as a next-generation gain medium. Nevertheless, further power/energy scaling and optimization cannot rely on empirical development alone. Spectroscopic data is necessary for modeling laser dynamics, energy storage, and thermal effects, all of which ultimately determine achievable laser performance. This section presents comprehensive characterization, including Raman spectra, absorption and emission cross sections, upper level lifetimes, low-temperature spectroscopy, and analysis of vibronic transitions, which will eventually guide the design of advanced Ho:CALGO laser systems.

### 3.1. Polarized Raman spectra

The factor group analysis [94] for the primitive cell of the $D^{17}_{4h}$ symmetry predicts the following set of irreducible representations at the center of the Brillouin zone (k = 0): $\Gamma = 2A_{1g} + 2E_g + 4A_{2u} + 5E_u + B_{2u}$, of which Raman-active modes are all even (gerade) species involving vibrations of mainly one type of atom: $A_{1g} + E_g$ for Ca|Gd atoms and $A_g + E_g$ for O; the IR-active modes are $3A_{2u} + 4E_u$; two modes ($A_{2u} + E_u$) are acoustic and one ($B_{2u}$) is silent.

The polarized Raman spectra of the ~3 at.% Ho:CALGO crystal are presented in Fig. 3, as measured using an *a*-cut sample. We use Porto's notation [97]: $m(nk)\bar{l}$, where m and l are the directions of propagation of the incident and scattered light, respectively (for the confocal geometry, $m \equiv \bar{l}$), and n and k are the polarization states of the incident and scattered light, respectively. Four spectra were measured, $a(nk)\bar{a}$, where n, k = π or σ polarization. The Raman spectra are strongly polarized.

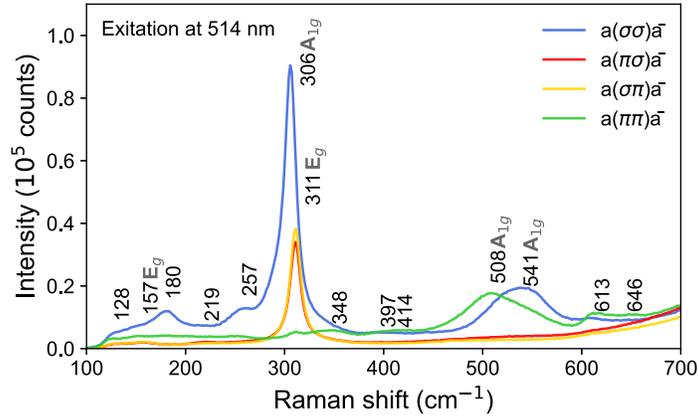

Fig. 3. Room temperature polarized Raman spectra of the ~3 at.% Ho:CALGO crystal in the $a(ij)\bar{a}$ geometry: $i, j$ = π, σ polarizations, $\lambda_{exc}$ = 514 nm, *numbers* indicate the Raman frequencies in cm$^{-1}$.

The $a(nn)\bar{a}$ geometry selects the $A_{1g}$ species, and in the $a(nk)\bar{a}$ geometry, the $E_g$ modes appear. For the latter geometry, as expected, two intense modes at ~157 and 311 cm$^{-1}$ appear. The low-frequency mode is assigned to Ca|Gd vibrations in the *a-b* plane and the high-frequency one – to the similar O vibrations. For the $a(\pi\pi)\bar{a}$ $A_{1g}$ geometry, the spectrum presents the most intense band at 306 cm$^{-1}$ and a weaker band at ~541 cm$^{-1}$. The former one is due to Ca|Gd vibrations along the *c*-axis and the band with a complex structure (clearly revealed for the $a(\sigma\sigma)\bar{a}$ geometry) at ~508/541 cm$^{-1}$ – to O vibrations. Thus, we identify all four Raman-active modes. The high-frequency bands at ~613 and 646 cm$^{-1}$ are due to defect-induced modes.

The low-phonon energy behavior of Ho:CALGO is expected to reduce the rates of multiphonon non-radiative relaxation from the $^5I_7$ and $^5I_6$ Ho$^{3+}$ manifolds, which is relevant for the operation of 2-μm and (potentially) 3-μm lasers.

### 3.2. Polarized absorption spectra and Judd-Ofelt analysis

The tetragonal CALGO crystal is optically uniaxial (the optical axis is parallel to the c-axis). Thus, there exist two principal light polarizations, $\boldsymbol{E} \parallel \boldsymbol{c}$ (π) and $\boldsymbol{E} \perp \boldsymbol{c}$ (σ) with the corresponding refractive indices $n_e > n_o$ (optically positive uniaxial crystal). The spectroscopic properties were measured using a-cut crystals giving access to both principal light polarizations. The overview absorption spectrum of a ~3 at.% Ho:CALGO crystal (Ho$^{3+}$ ion density: $N_{Ho}$ = 3.78×10$^{20}$ at/cm$^3$) for the two eigen polarization states, π and σ, is shown in Fig. 4. The attribution of electronic transitions is after Carnall *et al*. [98]. The spectra are strongly polarized, generally showing higher transition cross-sections for π-polarized light. In the UV, the sharp peaks arising from the host-forming Gd$^{3+}$ cations are visible [99]. The UV absorption edge is found at ~223 nm (optical bandgap: $E_{g,opt}$ = 5.56 eV), being consistent with that in the matrix. Ho:CALGO exhibits a relatively broad transparency range (among oxide crystals), and its IR absorption edge is found at 6.7 μm.

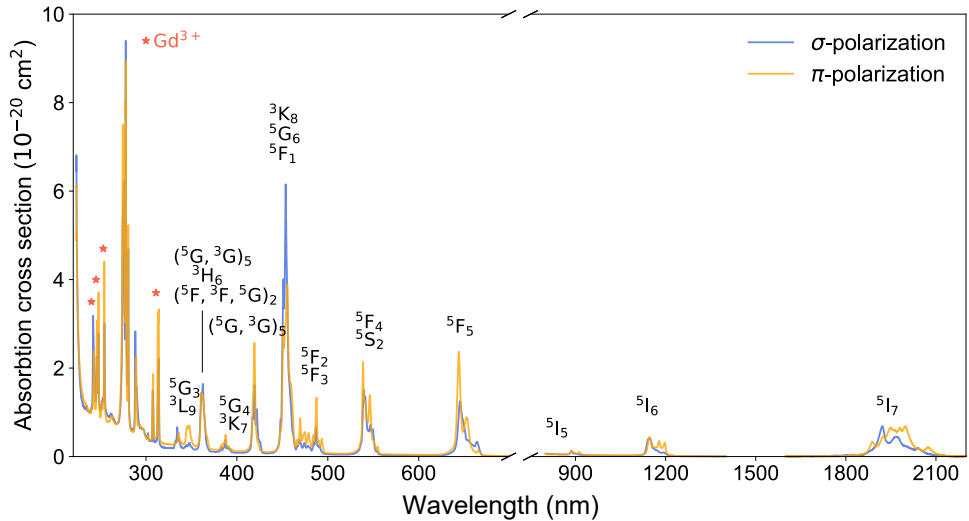

Fig. 4. Room temperature absorption cross-section, $\sigma_{abs}$, spectra of Ho$^{3+}$ ions in the CALGO crystal for σ and π polarized light. Asterisks mark the absorption lines of the host-forming Gd$^{3+}$ ions (not included in the calculation of cross-sections).

The Judd-Ofelt formalism [100] was applied to electric-dipole (ED) contributions to transition intensities. The magnetic-dipole (MD) contributions (for transitions obeying the selection rule $\Delta J = J - J' = 0, \pm 1$) were calculated separately within the Russell–Saunders approximation on wave functions of Ho$^{3+}$ assuming a free ion. The set of reduced squared matrix elements $U^{(k)}$ was taken from Ref. [101]. The dispersion curves of CALGO reported in Ref. [71] were used. All the values were considered as polarization-averaged, <..> = (2σ + π)/3.

The modified Judd-Ofelt (mJ-O) theory allows for better description of the experimental absorption oscillator strengths of $Ho^{3+}$ ions in CALGO. This parametrization scheme accounts for configuration interaction with the lower-energy excited configuration of the opposite parity ($4f^{n-1}5d^1 = 4f^95d^1$ for $Ho^{3+}$), for which the ED line strengths of the $J \rightarrow J'$ transitions $S^{ED}(JJ')$ are given by [102]:

$$S_{calc}^{ED}(JJ') = \sum_{k=2,4,6} U^{(k)} \tilde{\Omega}_k, \quad (1a)$$

$$\tilde{\Omega}_k = \Omega_k[1 + 2\alpha(E_J + E_{J'} - 2E_f^0)]. \quad (1b)$$

$$U^{(k)} = \langle (4f^m)SLJ \| U^k \| (4f^m)S'L'J' \rangle^2. \quad (1c)$$

Here, the intensity parameters $\tilde{\Omega}_k$ are the linear functions of energies ($E_J$ and $E_{J'}$) of the two multiplets involved in the transition, $E_f^0$ is the mean energy of the $4f^n$ configuration, $U^{(k)}$ are the reduced squared matrix elements, $\Omega_k$ are the intensity (J–O) parameters ($k$ = 2, 4, 6), and $\alpha$ is a phenomenological parameter determined by the covalency effects and interaction with the excited configuration of the opposite parity. The obtained intensity parameters are $\Omega_2$ = 11.221, $\Omega_4$ = 13.586 and $\Omega_6$ = 6.048 [$10^{-20}$ cm$^2$] and $\alpha$ = 0.070 [$10^{-4}$ cm] [103].

The probabilities of radiative spontaneous transitions for emission channels $J \rightarrow J'$ could then be determined from the corresponding line strengths in emission:

$$A_{calc}^{\Sigma}(JJ') = \frac{64\pi^4 e^2}{3h(2J+1)\langle\lambda\rangle^3} n \left(\frac{n^2+2}{3}\right)^2 S_{calc}^{ED}(JJ') + A_{calc}^{MD}(JJ'). \quad (2)$$

Using the $A_{calc}$ values, one can compute the total probabilities of spontaneous radiative transitions from a given excited state $A_{tot}$, its radiative lifetime $\tau_{rad}$ and the luminescence branching ratios for particular emission channels $\beta_{JJ'}$:

$$\tau_{rad} = \frac{1}{A_{tot}}, \text{ where } A_{tot} = \sum_{J'} A_{calc}^{\Sigma}(JJ'); \quad (3a)$$

$$B(JJ') = \frac{A_{calc}^{\Sigma}(JJ')}{A_{tot}}, \quad (3b)$$

The selected transition probabilities are summarized in Table 2. For the $^5I_7$ $Ho^{3+}$ multiplet, the mJ-O theory yields $\tau_{rad}$ = 3.32 ms. Considering the residual difference between the experimental and calculated absorption oscillator strengths for the $^5I_8 \rightarrow ^5I_7$ absorption transition, the upper limit for $\tau_{rad}(^5I_7)$ is 5.04 ms. This difference arises from polarization-dependent selection rules governing electronic transitions between specific Stark sublevels, as well as from the energy-level structure of the involved multiplets. The radiative lifetime for the $^5I_6$ manifold is 1.31 ms, and the luminescence branching ratio for the 2.9-μm emission is $\beta_{JJ'}$ = 13.7%.

**Table 2. Selected Probabilities of Spontaneous Radiative Transitions[a] of $Ho^{3+}$ Ion in CALGO Calculated Using the mJ-O Theory.**

| Emitting state | Terminal state | <λ>, nm | $A_{calc}^{\Sigma}(JJ')$, s$^{-1}$ | $\beta_{JJ'}$ | $A_{tot}$, s$^{-1}$ | $\tau_{rad}$, ms |
|---|---|---|---|---|---|---|
| $^5I_7$ | $^5I_8$ | 1935 | 273.0$^{ED}$+42.0$^{MD}$ | 1 | 301.5 | 3.32 |
| $^5I_6$ | $^5I_7$ | 2909 | 76.9$^{ED}$+20.4$^{MD}$ | 0.137 | 763.6 | 1.31 |
|  | $^5I_8$ | 1162 | 654.6$^{ED}$ | 0.863 |  |  |

*<λ$_{em}$> – mean emission wavelength of the emission band, $A_{calc}^{\Sigma}(JJ')$ – probability of radiative spontaneous transition, $\beta_{JJ'}$ – luminescence branching ratio, $A_{tot}$ and $\tau_{rad}$ – total probability of radiative spontaneous transitions and the radiative lifetime, respectively, ED and MD electric- and magnetic-dipole, respectively.

### 3.3. Polarized absorption and stimulated-emission cross-sections at 2 μm

Figure 5 depicts the absorption, $\sigma_{abs}$, and stimulated-emission (SE), $\sigma_{SE}$, cross-sections for the $^5I_8 \leftrightarrow ^5I_7$ transitions of $Ho^{3+}$ ions in CALGO around 2 μm, for two principal light polarizations, π and σ. The $^5I_8 \to {}^5I_7$ transition in absorption is used for in-band (resonant) pumping of $Ho^{3+}$ ions, *e.g.*, by means of Tm-fiber lasers. For π-polarized light, a broad plateau in the absorption spectrum spans from 1.90 to 1.95 μm, corresponding to $\sigma_{abs}$ of about $0.65\times10^{-20}$ $cm^2$, while for σ-polarization, the peak $\sigma_{abs}$ is slightly higher, measuring $0.69\times10^{-20}$ $cm^2$ at 1.92 μm. These wavelengths align well with the emission range of commercially available high-power Tm-fiber lasers. Nearly identical transition cross-sections for both eigen polarization states remove the need for polarized pumping, while the inhomogeneously broadened $Ho^{3+}$ absorption bands relax wavelength-stabilization requirements for in-band pumping, at the expense of reduced absorption compared to ordered crystals.

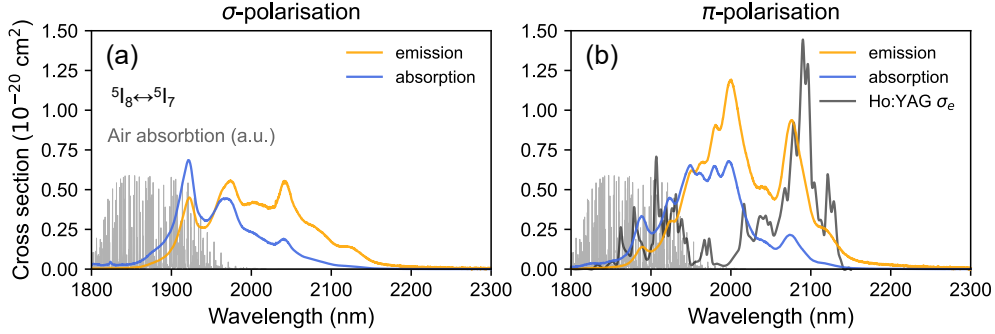

Fig. 5 Room temperature absorption, $\sigma_{abs}$, and stimulated-emission, $\sigma_{SE}$, cross-sections for the $^5I_8 \leftrightarrow {}^5I_7$ transitions of $Ho^{3+}$ ions in CALGO, light polarization: (a) σ and (b) π. In (b), the $\sigma_{SE}$ spectra for Ho:YAG are given for comparison.

The stimulated-emission cross-section, $\sigma_{SE}$, spectra around 2 μm were calculated using a combination of the Füchtbauer-Ladenburg (F-L) formula and the reciprocity method (McCumber equation). This approach removes the influence of reabsorption from the measured luminescence spectra and provides an independent consistency check. The F-L formula reads [104]:

$$\sigma_{SE}(\lambda) = \frac{\lambda^5}{8\pi<n>^2\tau_{rad}c1/3}\frac{\beta_{JJ'}W'_i(\lambda)}{\sum_{i=\pi,2\sigma}\int \lambda W'_i(\lambda)d\lambda}, \quad (4)$$

where, $\lambda$ is the light wavelength, $<n>$ is the polarization-averaged refractive index at $<\lambda_{em}>$, $\tau_{rad}$ and $\beta_{JJ'}$ are the radiative lifetime of the upper laser level and the luminescence branching ratio, respectively, and $W'(\lambda)$ is the luminescence spectrum corrected for the apparatus function of the set-up and the structured water absorption in the air, for the *i*-th light polarization (i = π or σ).

The reciprocity method is based on the following calculation [105]:

$$\sigma^i_{SE}(\lambda) = \sigma^i_{abs}(\lambda)\frac{Z_1}{Z_2}\exp\left(-\frac{(hc/\lambda)-E_{ZPL}}{kT}\right) \quad (5a)$$
$$Z_m = \sum_k g^m_k \exp(-E^m_k/kT). \quad (5b)$$

Here, $k$ is the Boltzmann constant, $T$ is the temperature, $E_{ZPL}$ is the energy of the zero-phonon-line (ZPL) transition between the lowest Stark sub-levels of the involved multiplets, $Z_m$ are the partition functions of the lower ($m = 1$) and upper ($m = 2$) manifold, and $g^m_k$ is the degeneracy of the Stark sub-level $k$ and energy $E^m_k$ relative to the lowest sub-level of each multiplet.

For this calculation, we used the experimental crystal-field splitting of the $^5I_7$ and $^5I_8$ $Ho^{3+}$ manifolds determined by low-temperature spectroscopy. The calculation of $\sigma_{SE}$ spectral profiles by the two complementary approaches provides an estimate for the radiative lifetime of the $^5I_7$ state of 5.7±0.5 ms, being slightly above the upper limit estimation of the Judd-Ofelt analysis. At the wavelengths where laser operation is expected, $\sigma_{SE}$ is $0.94\times10^{-20}$ $cm^2$ at 2076 nm and $0.25\times10^{-}$

$^{20}$ cm$^2$ at 2120 nm (for π-polarized light). The polarization anisotropy of stimulated-emission properties of Ho$^{3+}$ ions in *a*-cut CALGO crystals will imply laser operation on π-polarization.

It is worth comparing the SE spectral profiles of Ho$^{3+}$ ions in the disordered CALGO crystal with those in the ordered YAG compound, Fig. 5(b). Ho:CALGO features smooth, "glassy-like" emission spectra supporting the generation of sub-100 fs pulses.

### 3.4. *Luminescence lifetimes*

The dynamic of luminescence from the $^5I_7$ and $^5I_6$ Ho$^{3+}$ manifolds was further studied, revealing a nearly single exponential decay under nanosecond excitation, Fig. 6, suggesting the validity of a "quasi-center" model proposed by Alexander Kaminskii for structurally disordered crystals. For the measurements, the crystals were finely ground to reduce the effect of reabsorption on the measured lifetimes. For 1 at.% Ho$^{3+}$ doping at ambient temperature, they amount to $\tau_{lum}$ = 5.27 ms ($^5I_7$) and 0.237 ms ($^5I_6$). For the $^5I_7$ manifold, the measured luminescence lifetime is close to the estimates of the radiative one (the upper limit of the Judd-Ofelt analysis and the value extracted from the $\sigma_{SE}$ calculation), suggesting a luminescence quantum efficiency close to unity. This aligns well with the energy gap law, $\Delta E / h\nu_{ph,max} > 7$ (where $\Delta E$ is the energy gap to the highest sub-level of the lower-lying manifold), ruling out multiphonon non-radiative processes. On the contrary, the multiphonon path is significant for the $^5I_6$ state responsible for the 2.9-μm emission. As a reference, for a 1 at.% Ho$^{3+}$-doped YAG crystal with a higher effective phonon energy, the luminescence lifetimes are 8.0 ms and 44 μs. This variation reflects both the difference in the crystal-field strength and the vibronic properties of the material.

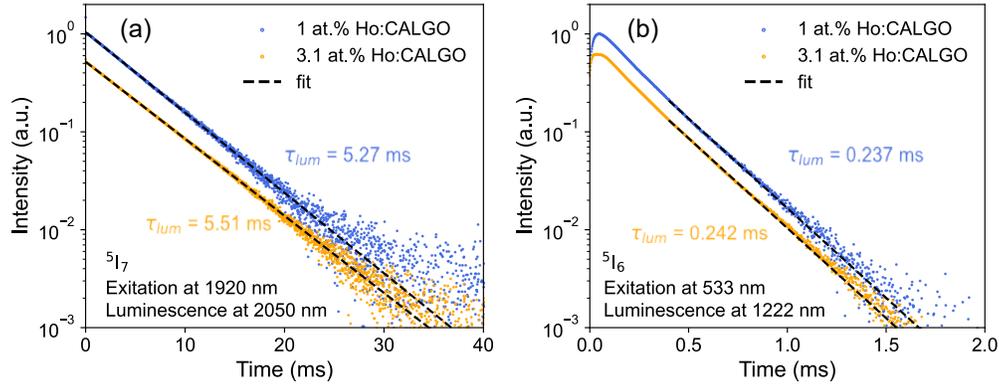

Fig. 6. Luminescence decay curves for 1 at.% and 3.1 at.% Ho:CALGO: (a) $^5I_7$ state and (b) $^5I_6$ state. *Points* – experimental data, *lines* – their single-exponential fits, $\lambda_{exc}$ and $\lambda_{lum}$ – excitation and detection wavelengths, respectively. The curves are measured at room temperature for finely powdered samples to avoid radiation trapping.

### 3.5. *Polarized gain spectral profiles at 2 μm*

To describe the spectral gain distribution in quasi-3-level lasers with intrinsic reabsorption from the terminal laser manifold, gain cross-sections, $\sigma_{gain} = \beta\sigma_{SE} - (1-\beta)\sigma_{abs}$, are usually calculated, where $\beta = N_2/N_{dop}$ is the inversion ratio, $N_1(^5I_8) + N_2(^5I_7) \approx N_{dop} = N_{Ho}$ (here the populations of other manifolds are neglected). Polarization-resolved gain spectral profiles allow one to assess the accessible gain bandwidth (which is particularly relevant for wavelength tunable and mode-locked operation), the polarization state selected by the gain competition in the absence of polarization-selective elements in an anisotropic crystal, and expected laser wavelengths in the free-running regime of operation. Two distinct operation regimes for Ho:CALGO lasers at 2.1 μm can be predicted:

i) a high-inversion regime (weaker reabsorption due to ground-state bleaching: $\beta > 0.25$) which is expected for low-doped (<1 at.% Ho) and relatively short (a few mm-thick) crystals providing smaller net gain. This regime supports laser emission at 2.08 μm in π-polarization with an available gain bandwidth (FWHM) of about 35 nm, Fig. 7(a,b). This regime of operation mainly relies on phonon-broadened electronic transitions of $Ho^{3+}$ ions;

ii) a low inversion regime (strong reabsorption, weak ground-state bleaching: $\beta < 0.2$) which can be observed for heavily doped (a few at.% Ho) and/or relatively long (cm-thick) crystals maximizing the pump absorption efficiency and gain, where the Ho:CALGO laser would tend to operate on the longest-wavelength electronic transition at ~2.12 μm ($5133(\Gamma_{3,4})$ cm$^{-1}$ → $438(\Gamma_2)$ cm$^{-1}$, see below) with a non-negligible contribution from the phonon-terminating mechanism, Fig.7(b). The available gain bandwidth is as broad as ~60 nm (for both polarizations), supporting the generation of much shorter pulses in the mode-locked regime, Fig. 7(c).

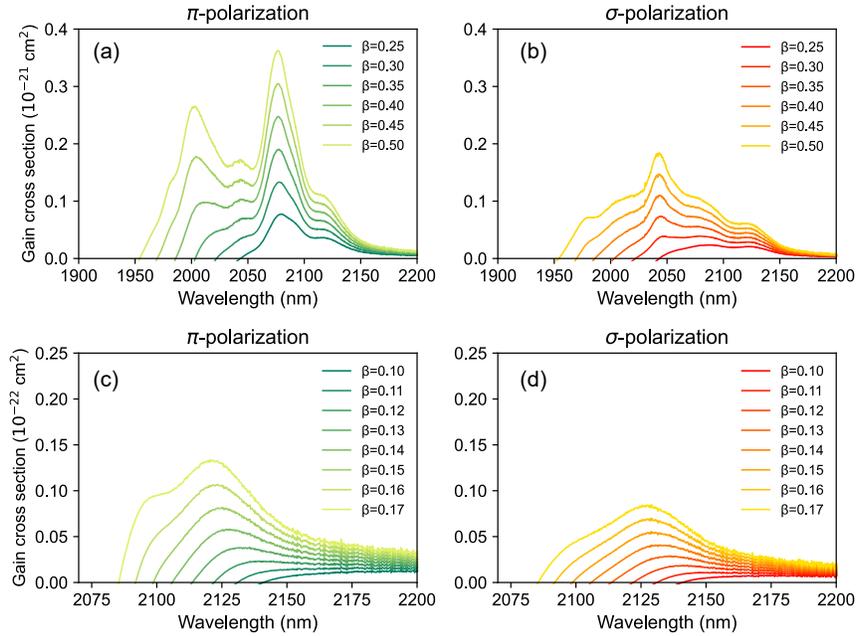

Fig .7. Room temperature gain spectral profiles of $Ho^{3+}$ ions in CALGO: (a,b) high-inversion regime, and (c, d) low-inversion regime.

### 3.6. *Polarized low temperature spectroscopy of $Ho^{3+}$ ions*

To determine the crystal-field (Stark) splitting of the three lowest-lying $Ho^{3+}$ manifolds relevant for operation of 2-μm and 3-μm lasers, low-temperature (LT, 12 K) polarized spectroscopy was employed, Fig. 8(a-c). Since $Ho^{3+}$ ions replace for the host-forming $Gd^{3+}$ cations occupying the $C_{4v}$ symmetry sites in CALGO, the total number of Stark sub-levels for a $^{2S+1}L_J$ multiplet with an integer $J$ is less than $2J + 1$, *i.e.*, it is 13, 11, and 10 for the three lower-lying states of $Ho^{3+}$ involved in the 2.1-μm and 2.9-μm transitions, namely $^5I_8$, $^5I_7$ and $^5I_6$, respectively. Following the theoretical data on the crystal-field splitting previously reported by Hutchinson *et al.* [106], for an isostructural Ho:CaYAlO$_4$ crystal obtained using a quantum mechanical point charge model, the assignment of electronic transitions was performed (see Table 3). The zero-phonon line (ZPL) energy for the $^5I_7 \rightarrow {}^5I_8$ transition is 5133 cm$^{-1}$ and the partition functions are $Z_1(^5I_8) = 8.315$ and $Z_2(^5I_7) = 8.966$ (at room temperature). The total Stark splitting of the ground-state,

$\Delta E(^5I_8)$, is relatively large, 438 cm$^{-1}$, which explains the broadband emission from Ho$^{3+}$ ions extending beyond 2.1 μm.

Remarkably, the absorption and emission spectra at 12 K, when the homogeneous phonon broadening is greatly suppressed, retain their broad nature, suggesting a strong inhomogeneous broadening arising from structural disorder. The second coordination sphere of Ca$^{2+}$ | RE$^{3+}$ (RE = Gd, Ho) cations in CALGO is composed of 9 nearest neighbor 4$e$ sites [cf. Fig. 8(d)]. The origin of the disorder is the second coordination sphere of RE$^{3+}$ ions, namely, the charge difference between the Ca$^{2+}$ and RE$^{3+}$ cations and the large difference in the metal-to-metal interatomic distances [107,108]. As only one apical site corresponds to the shortest Ca|RE – Ca|RE distance, it induces two families of local environment if it is occupied by Ca$^{2+}$ or RE$^{3+}$ cations. Each family contains surroundings with different distributions of Ca$^{2+}$ and RE$^{3+}$ cations over the eight remaining 4$e$ sites. This leads to significant inhomogeneous broadening of absorption and emission bands of optically active Ho$^{3+}$ cations. Despite this complex behavior, Ho$^{3+}$ ions in CALGO present a near single-exponential luminescence decay at low doping levels, suggesting the existence of an optical "quasi-center", a concept that generalizes the properties of several activator centers that slightly differ in structure but have very similar Stark splitting of energy states and rates of spontaneous transitions.

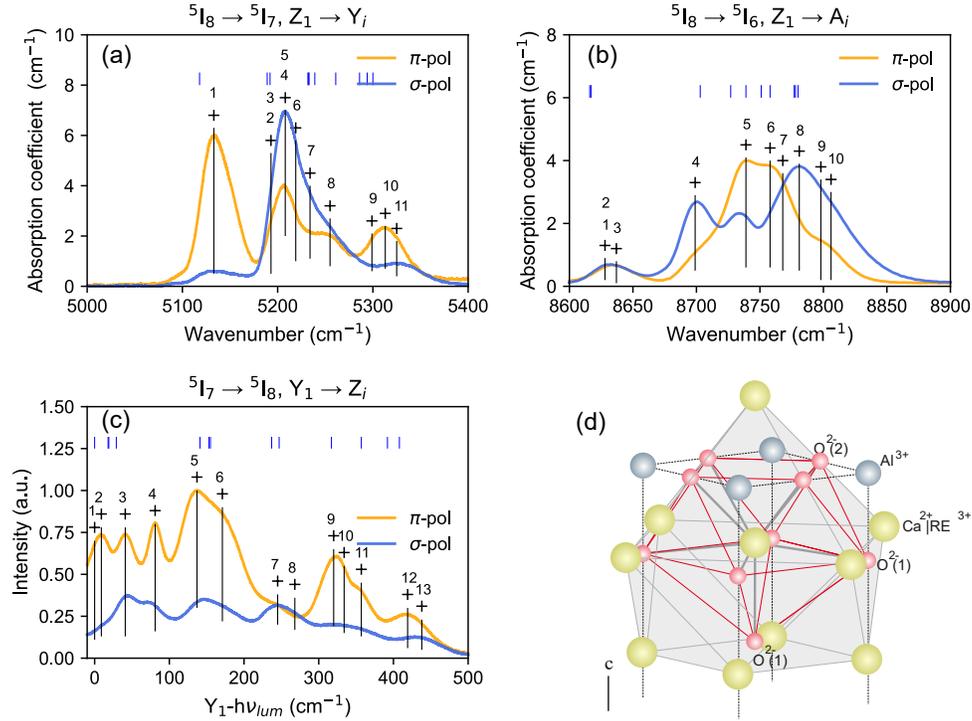

Fig. 8. (a-c) Low-temperature (12 K) spectroscopy of Ho$^{3+}$ ions in CALGO: (a,b) absorption spectra: (a) $^5I_8 \rightarrow {}^5I_7$ and (b) $^5I_8 \rightarrow {}^5I_7$; (c) luminescence spectra, $^5I_7 \rightarrow {}^5I_8$, "+" mark the assigned electronic transitions, using phenomenological notations for Stark sub-levels: $X_i$ $Y_j$, and $Z_k$ for the $^5I_6$, $^5I_7$ and $^5I_8$ manifolds, respectively, *vertical dashes* – theoretical crystal field splitting for Ho:CALYO after Hutchinson *et al.* [106]. (d) Second coordination sphere of Ho$^{3+}$ ions by Ca$^{2+}$|Gd$^{3+}$ cations in CALGO.

Table 3. Experimental Crystal-Field Splitting for Ho$^{3+}$ Ions in CALGO.

| $^{2S+1}L_J$ | Total no. | Energy (irreducible representation Γ), cm$^{-1}$ |
| --- | --- | --- |
| $^5I_8$ | 13 | 0($\Gamma_{3,4}$); 9($\Gamma_2$); 41($\Gamma_1$); 81($\Gamma_1$); 137($\Gamma_2$); 171($\Gamma_{3,4}$); 245($\Gamma_1$); 268($\Gamma_2$); 320($\Gamma_{3,4}$); 334($\Gamma_1$); 357($\Gamma_1$); 419($\Gamma_{3,4}$); 438($\Gamma_2$) |
| $^5I_7$ | 11 | 5133($\Gamma_{3,4}$); 5193(2$\Gamma_2$); 5208($\Gamma_{3,4}$,$\Gamma_1$); 5219($\Gamma_2$); 5234($\Gamma_{3,4}$); 5255($\Gamma_1$); 5299($\Gamma_1$); 5313($\Gamma_{3,4}$); 5325($\Gamma_2$) |
| $^5I_6$ | 10 | 8628(2$\Gamma_2$); 8637($\Gamma_{3,4}$); 8699($\Gamma_1$); 8739($\Gamma_1$); 8758($\Gamma_{3,4}$); 8768($\Gamma_2$); 8781($\Gamma_2$); 8798($\Gamma_1$); 8806($\Gamma_{3,4}$) |

### 3.7. *Multiphonon-assisted long-wave emission*

The luminescence spectra of Ho$^{3+}$ ions in CALGO extend well beyond 2 μm, a property which cannot be solely explained by the crystal-field splitting if the manifolds involved in the $^5I_7 \rightarrow$ $^5I_8$ electronic transition and inhomogeneous spectral line broadening. Ho$^{3+}$ ions, located at the end of the lanthanide series, are known to experience a strong electron-phonon interaction leading to the appearance of exponential multi-phonon sidebands in the long-wave part of their absorption and emission spectra. Figure 9(a) depicts the long-wave $\sigma_{SE}$ and $\sigma_{abs}$ spectral profiles of Ho$^{3+}$ ions up to 2.35 μm, plotted in a semi-log scale. As pointed out above, the longest wavelength purely electronic transition of Ho$^{3+}$ ions, 5133($\Gamma_{3,4}$) cm$^{-1}$ → 438($\Gamma_2$) cm$^{-1}$, occurs at 2130 nm. Emission beyond this wavelength originates from phonon-assisted (Stokes) transitions to vibronic states. These states can be constructed by adding the host-lattice phonon energies to the highest-lying Stark sublevel of the ground manifold, Fig. 9(b), which suggests emission extending to 2.3 μm for single-phonon processes. In practice, multiple lattice vibrations from the phonon bath – including both Raman- and IR-active modes – can participate in emission, producing a smooth, nearly exponential phonon sideband.

Auzel proposed expressions for the dependence of the multiphonon non-radiative relaxation rates of the Stokes (S) and anti-Stokes (AS) phonon-assisted processes on the transition energy $E$ [109]. Braud *et al.* applied these formulas for describing the phonon sidebands of the transition cross-section spectra [110]:

$$\sigma_S = \sigma_0 \exp(-\alpha_S \Delta E), \tag{4a}$$

$$\sigma_{AS} = \sigma_0 \exp(-\alpha_{AS} \Delta E), \tag{4b}$$

where $\Delta E$ is the energy mismatch between the vibronic and purely electronic transitions, $\sigma_0$ is the cross-section at the photon energy of the electronic transition, and $\alpha_S$ and $\alpha_{AS}$ are host-dependent factors. This exponential behavior reflects a weak-coupling ion–lattice interaction, typical of $4f^n$ transitions. The material factors can be expressed as:

$$\alpha_S = (h\nu_{ph})^{-1}\{\ln[<N>/(S_0(<n>+1))]-1\}, \tag{5a}$$

$$\alpha_{AS} = \alpha_S + 1/(kT), \tag{5b}$$

where, $h\nu_{ph}$ is the maximum energy of the phonon bath of the host matrix, $<N>$ is the average number of phonons involved in the phonon-assisted transition, $S_0$ is the Pekar-Huang-Rhys coupling constant and $<n>$ is the occupation number for the effective phonons of energy $h\nu_{ph}$. For the Ho:CALGO crystal, the best-fit $\alpha_S$ and $\alpha_{AS}$ values are 7±0.5×10$^{-3}$ cm and 12±0.5×10$^{-3}$ cm, respectively. Long-wave multiphonon-assisted emission of Ho$^{3+}$ ions can extend their gain to longer wavelengths, which, in combination with the use of a structurally disordered host matrix such as CALGO featuring a significant inhomogeneous spectral broadening, is expected to support the generation of few-optical-cycle pulses, *e.g.*, from Kerr-lens mode-locked lasers.

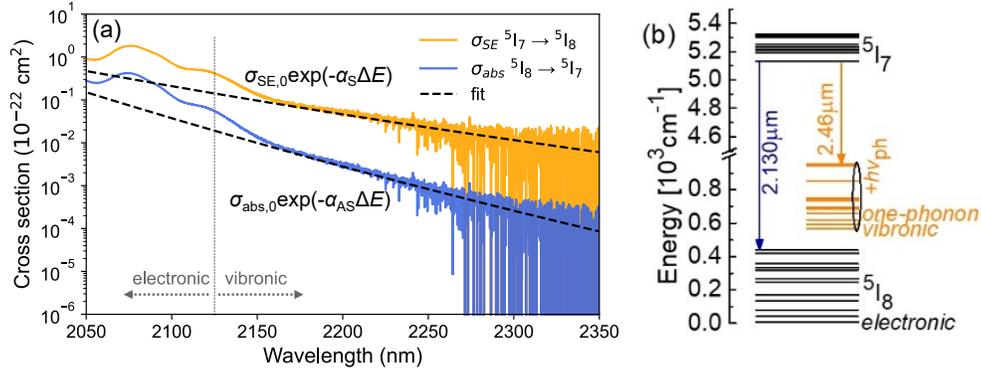

Fig. 9. (a) Phonon-terminated absorption (anti-Stokes) and emission (Stokes) of Ho:CALGO beyond 2 μm for π-polarized light; (b) The scheme of electronic (*black*) and one-phonon vibronic (*orange*) energy level relevant for understanding the long-wave emission from $Ho^{3+}$ ions.

## 4. Mode-locked laser performance of Ho:CALGO

### 4.1 Continuous wave laser operation

Continuous wave (CW) laser characteristics were first investigated using a Brewster cut 15-mm Ho(3.1 at.%):CALGO crystal in π-polarization geometry (E||c) and a X-shaped cavity for astigmatism compensation [89]. The gain crystal was pumped by a 1940-nm single-mode Tm-doped fiber laser. Output powers and laser spectra with 1% to 10% transmission output couplers (OC) are shown in Fig. 10. The slope efficiency kept increasing as OC transmission increased, and the highest slope efficiency of 52.4% was obtained with a 10% OC, and optical-to-optical (O-to-O) conversion efficiency reached 47% with a 7% OC. The increase in output power was not saturated at the maximum absorbed pump power of 22.8 W, and a drop in O-to-O efficiency was not observed, indicating excellent thermal conductivity of the crystal and the possibility of further power scaling. Here, the 52.4% slope efficiency is far below the Stokes limit of 91.5%.

The limited efficiency compared to the theoretical Stokes efficiency can be attributed to the energy transfer upconversion (ETU) process, which is a nonradiative ion–ion interaction between two neighboring ions in the upper laser level and is a common limitation to increase doping levels in Ho lasers. In the ETU process, one excited ion transfers its energy to a second excited ion, and excites it to a further upper level while the donor ion relaxes to the ground (or a lower-lying) state. As a result, two ions initially contributing to the population inversion are converted into a single ion in a higher excited state that typically relaxes rapidly via multiphonon decay. Consequently, ETU effectively depopulates the upper laser level, reducing the population inversion and optical gain. The strength of the ETU is defined by the average distance between ions in the upper laser level. For example, in conventional Ho:YAG lasers, the doping concentration is commonly chosen below 1%, i.e., ion density of less than $1.39 \times 10^{20}$ $cm^{-3}$, thus, the 3.1 at.% doping concentration (ion density of $3.85 \times 10^{20}$ $cm^{-3}$) would cause ETU more efficiently, limiting laser performance [17,63]. For the improvement of laser efficiency, investigation of ETU probability using spectroscopy and modeling of the laser dynamics would enable us to optimize the doping concentration and achieve higher laser efficiency. The optical spectra of the laser output are shown in Fig. 10(c). The typical operating wavelength of the CW oscillator is from 2120 nm to 2135 nm. Since the 2.1-μm Ho laser transition is a quasi-three-level system, the gain peak wavelength shifts toward blue as the inversion level increases, as shown in Fig. 10(b). Consequently, the laser spectra followed the behavior and a blue shift was observed with increasing output coupling loss.

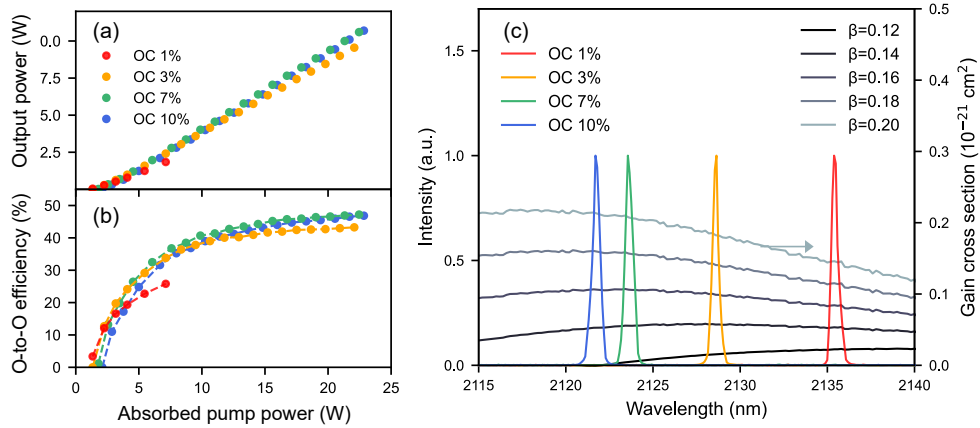

Fig. 10. CW laser output properties of Ho(3.1 at.%):CALGO laser. (a) output power as a function of absorbed pump power, (b) optical-to-optical conversion efficiency, and (c) optical spectra for different OCs.

*4.2 SESAM mode-locked laser oscillator*

The first attempt at ultrashort pulse generation with Ho:CALGO was demonstrated by SESAM mode-locking in 2022 [89]. The setup is shown in Fig. 11(a). As a gain medium, Ho(3.1 at.%):CALGO was used in π-polarization geometry (E∥c), which was pumped by a 1940-nm Tm-fiber laser. For mode-locking, a commercial SESAM was utilized. The measurement of the nonlinear reflectivity of the SESAM was performed using the 2.1-µm laser in the same paper, revealing the detailed specs and supporting the exploration of the limitation of mode-locked performance.

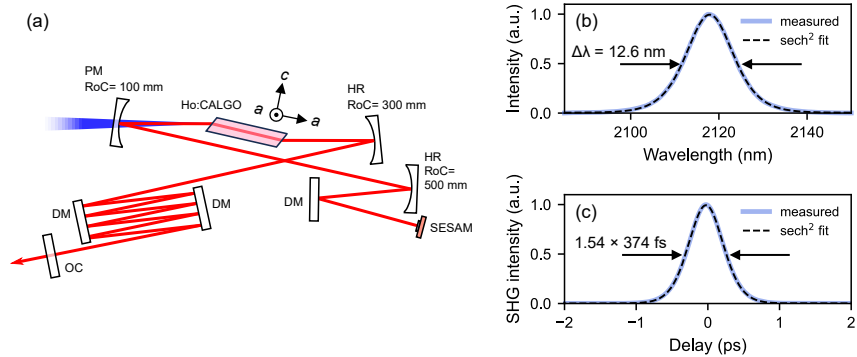

Fig. 11. (a)Experimental setup of SESAM mode-locked Ho:CALGO laser (PM: pump mirror, HR: highly reflective mirror, RoC: Radius of curvature, DM: dispersive mirror). (b) Optical spectrum and (c) autocorrelation trace of the SESAM mode-locked operation at 8.7-W average output power.

The maximum average power of 8.7 W was obtained under mode-locked operation with a 10% OC at a repetition rate of 84.4 MHz. The mode-locked laser spectrum and autocorrelation trace are shown in Fig. 11(b, c). The center wavelength of the soliton-shaped spectrum was 2117.9 nm, which is consistent with CW operation at a comparable level of loss. The spectral bandwidth of 12.6 nm provides the Fourier transform-limited pulse duration of 374 fs, and the pulse duration was measured to be 369 fs. The slightly shorter pulse duration than that of the Fourier limited value can be attributed to the slight deviation of actual laser pulses from the perfect soliton-shaped pulses. By utilizing smaller transmission OCs, broader bandwidths and shorter pulse durations were available at the expense of output powers. The shortest pulse duration of 306 fs was obtained using a 1% OC, delivering 1-W average output power with a red-shifted spectrum centered at 2133.8 nm and a spectral bandwidth of 15.1 nm. However, further pulse shortening was limited by a low modulation depth of the SESAM, making it difficult to suppress the CW background when the spectrum was further broadened. Nevertheless, the average output power reached the highest value among mode-locked bulk lasers in the 2-μm wavelength range. Moreover, further power scaling and pulse shortening should be possible by improving the SESAM parameters to obtain a higher damage threshold and larger modulation depth [111].

*4.3 Kerr-lens mode-locked laser oscillator*

KLM is a promising way to obtain shorter pulses, overcoming the limitation of the modulation depth of physical saturable absorbers, which can be seen in Section 3.2, and it also offers a fast response time, which creates a shuttering effect to further shorten the pulses. The first KLM Ho:CALGO laser was demonstrated in [90]. Since KLM requires control of the laser mode in both linear and nonlinear regimes, the laser cavity was carefully designed using the ABCD matrix method, and the Tm-doped fiber laser pump had a single transversal mode to ensure mode-matching for soft-aperture KLM along the >10 mm samples, which is impossible for most of the diode pumping schemes. In this case, σ-polarization (E⊥$c$) geometry was chosen as σ-polarization exhibits a flatter gain profile as shown in Fig. 7(d).

The careful cavity design and optimization of laser parameters enabled the Kerr-lens mode-locking of Ho:CALGO laser. The mode-locked spectrum was centered at 2135 nm with a spectral bandwidth of 45.3 nm. The pulse duration was 112 fs, achieving a pulse duration 3 times shorter than that of the SESAM mode-locked laser. Remarkably, this laser achieved 2-W average output power with a 5% OC, showing a high-power level among 100-fs class mode-locked lasers in the 2 μm range. The lower power compared to the SESAM mode-locked case can be rationalized by the use of a lower OC transmission compared to the SESAM mode-locked cavity. In the KLM experiments, when the OC transmission was increased to 7%, continuous-wave emission at shorter wavelengths around 2080 nm appeared in the laser spectrum, which was difficult to suppress due to the increased cavity loss, effectively constraining further power scaling via higher transmission. Nevertheless, further increase of output power is expected by enlarging the mode size in the gain crystal with the 5% OC, which would allow higher output power without compromising the pulse duration [13], at the expense of a higher mode-locking threshold.

The results mentioned above were often seeking to maximize output power, rather to push the pulse duration to the possible limits of the material. In more recent preliminary experiments, shorter pulse generation was explored in an attempt to exploit the widest possible bandwidth available [92]. To enhance self-phase modulation (SPM) to support further spectral broadening, the beam size on the gain crystal was set to a smaller size, and a lower transmission OC of 1% was used to increase intracavity power.

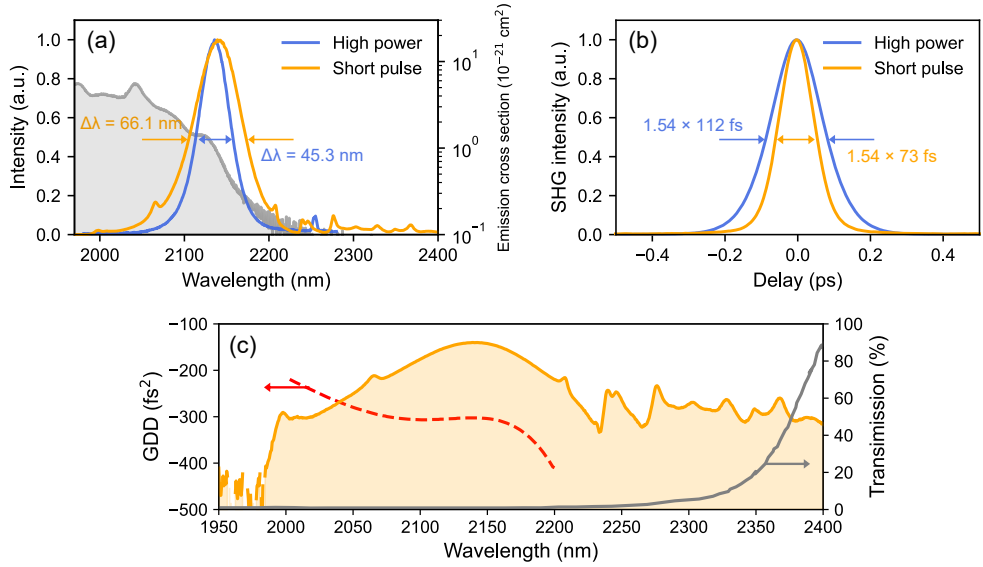

Fig. 12. (a) Optical spectra (blue and orange) and emission cross section spectrum (grey) in a logarithmic scale, and (b) autocorrelation traces of KLM Ho:CALGO lasers. (c) GDD profile of the chirped mirror (red, dashed) and transmission curve of the output coupler (solid grey) used in the short pulse KLM laser cavity, with the output spectrum (orange) in logarithmic scale for comparison.

As a result, pulses as short as 73 fs were obtained with an average output power of 170 mW. This marks the shortest pulse duration among ever-reported Ho-based mode-locked laser systems, albeit at a significant loss in power. The autocorrelation trace and mode-locked spectrum are shown in Fig. 12(a, b). The broader spectral bandwidth of 66.1 nm was obtained at a center wavelength of 2139 nm. Table 4 summarizes the mode-locked parameters obtained in both SESAM and KLM. The calculated phase shift shows a larger value for shorter pulse duration, which explains that SPM contributed significantly to pulse shortening. Further pulse shortening was limited by the bandwidth of the mirror coating. As shown in Fig. 12(c), subpeaks appeared on the shortest pulse spectrum around 2200-2400 nm, where the GDD of the mirror coating is no longer well controlled and not specified by the supplier. In this spectral region, the transmission of the OC starts to increase, making these subpeaks more prominent in the output spectrum. Further spectral broadening would be possible with other mirrors. The relatively low average output power compared to the other mode-locked results is primarily due to the use of a lower-transmission OC to minimize intracavity losses. Reduced cavity loss is essential for short-pulse generation, as it allows laser operation at lower inversion levels, where a smoother and broader effective gain bandwidth is available. Therefore, this result reflects a typical trade-off between pulse duration and output power in lasers based on gain media with finite bandwidth and inversion-dependent gain profiles.

Table 4. Summary of Ho:CALGO mode-locked laser output.

| Parameters | SESAM | KLM (high power) | KLM (short pulse) |
|---|---|---|---|
| $f_{rep}$ (MHz) | 84.4 | 91 | 89.6 |
| $P_{ic}$ (W) | 87 | 40 | 17 |
| $P_{out}$ (W) | 8.7 | 2 | 0.17 |
| $\lambda$ (nm) | 2117.9 | 2136 | 2139 |
| $\Delta\lambda$ (nm) | 12.6 | 45.3 | 66 |
| $\Delta\tau$ (fs) | 369 | 112 | 73 |
| GDD (fs$^2$) | -16800 | -2100 | -3300 |
| Phase shift (rad) | 0.71 | 1.08 | 2.37 |

Subsequent studies of the KLM Ho:CALGO laser led to the demonstration of a GHz repetition rate laser system [91]. Figure 13(a, b) shows the KLM cavity, and a bow-tie cavity was utilized with a 50-mm telescope to create a small focal spot, allowing for sufficient Kerr effect with a low pulse energy level. In mode-locked operation, unidirectional laser output was observed due to different strengths of Kerr-lensing for each directions, and the favorable oscillation direction that experiences larger gain is naturally selected. The direction can be controlled by changing the cavity alignment, however, output characteristics are consistent for both directional outputs. The output characteristics of the counterclockwise direction output are shown in Fig. 13(c, d). Soliton mode-locking was obtained with 1.68-W average power, and the spectrum was centered at 2159.1 nm with a spectral bandwidth of 55 nm. The pulse duration was 93 fs with a TBP of 0.329. The radio frequency measurement showed the stable single-pulse mode-locking at a fundamental repetition rate of 1.179GHz. We note that this laser system was very sensitive to alignment due to operation in very narrow stability regions, and thus requires engineering efforts to be taken to applications.

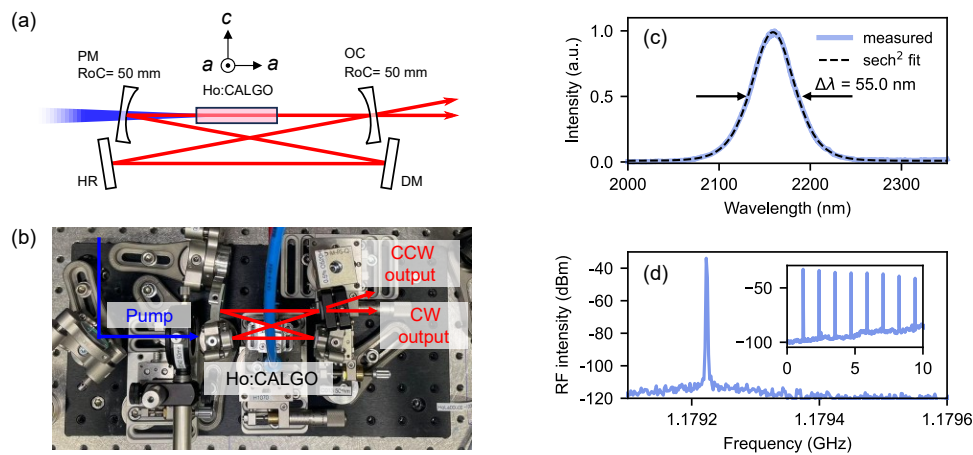

Fig. 13. (a) Experimental setup of the GHz repetition rate Ho:CALGO KLM laser. (b) Photo of the setup. Laser output properties of the counterclockwise oscillation, (c) optical spectrum and (d) radio frequency spectrum of the fundamental repetition rate (Inset: spanning up to 10 GHz).

*Potential future developments using mode-locked Ho:CALGO oscillators*

The high-power and broadband emission of Ho:CALGO lasers make them highly attractive for spectroscopic applications, in particular frequency comb spectroscopy for the high repetition rate oscillators. One attractive alternative is dual-comb spectroscopy enabling fast acquisition and high spectral resolution by simply measuring the interferogram of two combs. Recently, a single-cavity dual-comb Ho:CALGO laser was reported [112]. The first results showed watt-class average output power and 100-fs pulse duration for each comb, exploiting the full potential of Ho:CALGO. It is expected to enable high-dynamic-range spectroscopic measurements in future applications. Furthermore, for other single-comb precision spectroscopic applications, stabilization of the laser frequency is essential. In particular, detection and control of the carrier–envelope offset frequency ($f_{CEO}$) is a key requirement for full frequency-comb stabilization. Conventional $f$–$2f$ interferometry can be employed for $f_{CEO}$ detection; however, this requires octave-spanning spectral broadening. As a preliminary step toward $f_{CEO}$ detection and future stabilization, supercontinuum generation was demonstrated using a $TiO_2$ waveguide [113]. Semiconductor-based passive waveguides are particularly well suited for high-repetition-rate laser systems, as they enable efficient spectral broadening with small pulse energy levels. Using the generated supercontinuum, $f_{CEO}$ detection was successfully demonstrated, representing a first step toward comprehensive frequency-comb characterization and the realization of a fully stabilized dual-comb Ho:CALGO laser system.

## 5. Chirped pulse amplifier based on Ho:CALGO

The broad gain bandwidth is advantageous not only for generating ultrashort pulses via mode-locked lasers, but also for amplifying them, since high-gain amplifiers suffer from the gain narrowing effect due to the finite gain bandwidth of active media. Currently, Ho:YLF is a main workhorse for ultrafast laser amplifiers at 2.1 µm due to its large gain cross section, long upper-level lifetime of 16 ms [29], as well as excellent thermo-mechanical and thermo-optic properties. However, its narrow and structured gain bandwidth limits the available pulse duration to >2 ps [67,114]. In contrast, Ho:CALGO offers a much broader and smoother gain profile with moderate gain cross section, enabling femtosecond pulse amplification without spectral shaping techniques. The first Ho:CALGO CPA was demonstrated in [93]. A critical requirement for efficient amplification is a good spectral overlap between the seeder and the gain peak of the active medium. Depending on the inversion level, Ho:CALGO laser oscillators may therefore not be ideally suited since the peak gain wavelength of Ho:CALGO changes significantly with the inversion level. Mode-locked Ho:CALGO oscillators operating at a low inversion level emit at around 2120 nm, the gain of Ho:CALGO amplifiers at higher inversion is maximized at a shorter wavelength of 2078 nm for π-polarization, as shown in Fig. 7(a). This spectral mismatch leads to reduced amplification efficiency. In conventional Ho-doped laser amplifiers, wavelength-shifted sources such as frequency-shifted Er lasers [115,116] or an OPA driven by a high-power Yb-laser [67,114] were utilized as a seeder. These systems have the advantage of wavelength tunability, allowing one to obtain an ideal wavelength that matches the gain spectrum of the gain medium, however, they are rather complex and cumbersome. In this regard, Tm-doped or Tm,Ho-codoped lasers can be an attractive alternative. These operate around 2050-2100 nm depending on the material, and mode-locked techniques with these lasers are well established. In addition, the availability of 0.79-µm diode pumping enables a small and compact system. In the first demonstration of a Ho:CALGO CPA, a diode-pumped SESAM mode-locked Tm,Ho:CLNGG laser emitting at 2093 nm was utilized as a seeder [117]. Such sources offer a good balance between spectral compatibility, system simplicity, and performance, making them well-suited for Ho-doped laser amplifier systems.

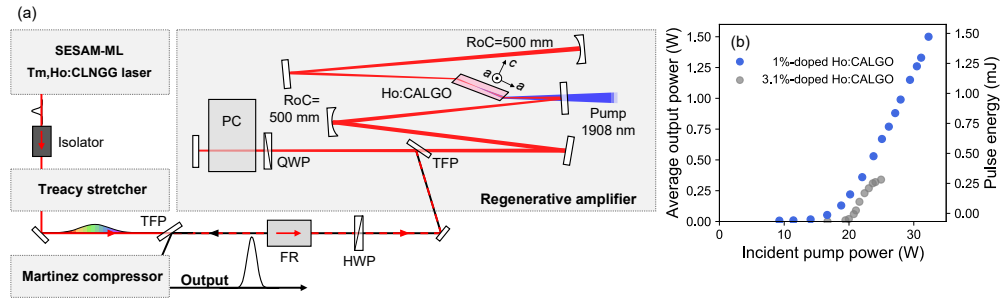

Fig. 14. (a) Experimental setup of the Ho:CALGO CPA (TFP: thin-film polarizer, FR: Faraday rotator, HWP/QWP, Half-/Quarter-wave plate, PC: Pockels cell). (b) Average power and pulse energy of amplifier output as a function of incident pump power with 1%- and 3.1%-doped Ho:CALGO.

The schematic of the Ho:CALGO laser amplifier is illustrated in Fig. 14(a). The system follows a conventional CPA arrangement, consisting of a seeder, a grating stretcher, a RA, and a grating compressor. Here, the doping concentration of the gain crystal plays a crucial role in determining the amplifier performance because of the pronounced ETU process. Since amplifiers operate at significantly higher inversion levels compared to oscillators, the probability of ETU is higher, which leads to a reduction of gain and additional heat generation. To evaluate the impact of doping concentration, the amplifier performance was compared using 3.1-at.% and 1-at.% doped Ho:CALGO crystals in π-polarization geometry. The length of each crystal was 14.48 mm and 26.42 mm for 3.1 at.% and 1 at.% doped crystal, respectively. Figure 14(b) presents the output power of the RA at a 1-kHz repetition rate. While the 3.1-at.% doped crystal exhibited a higher threshold and reached output power saturation around 25-W incident pump power, the 1-at.% doped crystal enabled higher energy extraction, reaching 1.5-W average power corresponding to 1.5-mJ pulse energy. This comparison shows the detrimental influence of ETU at higher doping concentrations, and demonstrates that reducing the doping concentration is a straightforward way to suppress ETU and improve amplifier performance.

Given the clear performance advantage of the lower-doped crystal, high-average-power operation was demonstrated with 1-at.% doped Ho:CALGO at a 100-kHz repetition rate. 11.2-W average output power and 112-µJ pulse energy were obtained at 28 roundtrips (RTs), and 9.3 W and 93 µJ were available after the Martinez compressor, i.e., 83% throughput. Figure 15(a) shows a 30-minute measurement of amplified pulse energy of the amplifier output and the compressor output. It shows good output stability with a root-mean-square (RMS) of less than 0.39% with a good beam quality of less than 1.2 for both axes. The optical spectrum of the amplifier output is shown in Fig. 15(b). The center wavelength was blue-shifted toward the gain peak of the Ho:CALGO, however, the amplified pulse maintained a broad bandwidth of 11 nm under the gain factor of $10^{5.4}$. That resulted in a compressed pulse duration of 750 fs and a peak power of 107 MW, while the Fourier transform limited (FTL) pulse duration was 626 fs.

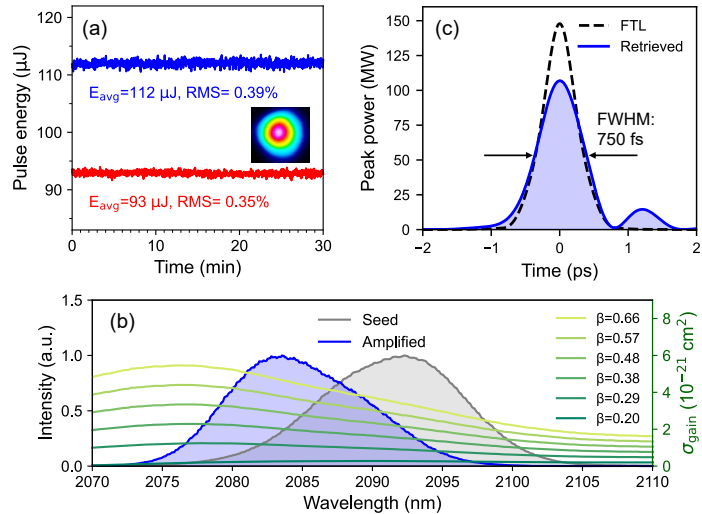

Fig. 15. (a) 30-minute operation of the Ho:CALGO CPA at 100 kHz (blue: amplifier output, red: compressor output). Inset: amplified beam profile after the compressor. (b) Optical spectra of the seed and amplified pulse and the gain cross sections of Ho:CALGO for π-polarization. (c) Temporal profile of the compressed pulse retrieved from frequency-resolved optical gating (FROG) measurement and FTL pulse calculated from the optical spectrum.

High-energy sub-picosecond pulses are attractive for nonlinear pulse compression, which enables the compression of pulses down to sub-100-fs with a simple single-stage compression stage. In [93], a multi-pass cell (MPC) compression was implemented to shorten the Ho:CALGO CPA output while preserving beam quality. The MPC provided a high transmission of ~90%, yielding 7.2-W average power and 72-µJ pulse energy with 8-W and 80-µJ incident power, and after 13 RTs, spectral broadening was obtained from 2.0 to 2.15 µm. Temporal compression was done using the material dispersion of YAG plates, resulting in 97-fs pulses, which is close to the 91-fs FTL pulse duration. The corresponding peak power reached 525 MW, representing a ~5× enhancement while maintaining excellent spatial quality ($M^2 < 1.1$). The high intensity would be attractive for strong field applications such as plasma-based THz generation.

## 6. Future Prospects of Ho:CALGO lasers

*6.1 Evaluation of performance limits in bulk geometry*

Further scaling of Ho:CALGO laser systems and optimization of their performance will require a more detailed understanding of both material properties and laser dynamics. As discussed in the CW laser and amplifier sections, ETU represents a key factor limiting laser and amplifier efficiency. Accurate determination of ETU probabilities through spectroscopic studies is therefore essential. Incorporating these measurements into comprehensive laser and amplifier models will allow optimization of doping concentrations, improving efficiency while mitigating thermal load at the same time.

In addition, characterization of the thermal properties of CALGO under $Ho^{3+}$-doping is crucial to explore the power scaling limit. While thermo-mechanical and thermo-optic properties have been reported for undoped and Yb-doped CALGO [39,71], these parameters would differ in the $Ho^{3+}$-doped system and its doping level. Systematic studies on the thermal properties of Ho:CALGO, including thermal conductivity, expansion, and thermo-optic coefficient, are therefore necessary to perform practical thermal simulations and laser design.

By combining comprehensive spectroscopic characterization with thermal and laser-dynamics modeling, it will be possible to explore the ultimate performance limits of Ho:CALGO-based laser systems in bulk geometry. Moreover, the quality of the grown crystals—including lattice defects, internal stress, and homogeneity—strongly influences both spectroscopic and thermal properties. Consequently, iterative improvements in crystal growth will remain essential efforts to achieve high-power, high-energy operation.

*6.2 Pulse duration scaling in mode-locked oscillators*

Further scaling of pulse duration in mode-locked Ho:CALGO oscillators is promising, however, technical challenges remain. Spectroscopic studies, shown in Section 3.5, indicate that Ho:CALGO offers an effective gain bandwidth of ~60 nm around 2120 nm. From this spectral bandwidth, a Fourier-transform-limited pulse duration can be calculated to be about 78 fs. However, shorter pulses than the gain bandwidth can be obtained, as long as a wider bandwidth can be generated, for example with SPM, and this bandwidth does not suffer reabsorption, i.e., a sufficiently broad bandwidth is operated at transparency. The measured mode-locked spectrum showed the generation of spectral components extending beyond the main gain bandwidth toward longer wavelengths. These components would be generated by strong intracavity SPM and by contributions of gain at longer wavelengths associated with multiphonon-assisted transitions, discussed in Section 3.7. In the current cavity configuration, the effective use of these extended spectral components is limited by the finite bandwidth and dispersion characteristics of the chirped mirror coating, as shown in Fig. 12(c). As a result, their spectral phase cannot be adequately controlled, preventing a contribution to further pulse shortening.

If dispersion compensation could be extended to properly manage these longer-wavelength components, the gain-bandwidth limit could be overcome, allowing pulse durations below 60 fs. Pulse-duration scaling using such behavior has been observed in ultrashort-pulse Yb- or Tm-based oscillators [62,73], where strong nonlinear spectral broadening combined with broadband dispersion control, as well as proper extraction of spectral components using a designed output coupler, enables operation beyond the conventional gain bandwidth.

On the short-wavelength side, spectral extension is typically constrained by the dichroic coating of the pump incoupling mirror. Since the pump and laser wavelengths in Ho-based systems are close, the short-wavelength edge of the high-reflectivity band is inherently limited. This constraint is clearly shown in the mode-locked spectrum, which exhibits a hard cut at the short wavelength edge, as shown in Fig. 12(c). In addition, near the edge of the coating bandwidth, the GDD curve typically starts to oscillate, which also hinders dispersion management over a broad spectral range. To overcome this limitation, alternative pumping and cavity concepts are required. In particular, cross-polarization pumping, as proposed in [118], decouples pump injection from the laser polarization and relaxes the coating constraints, thereby offering a viable route toward broader usable bandwidth and shorter pulse durations.

### 6.3 Pulse energy scaling in bulk amplifiers

In the present work, a bulk Ho:CALGO amplifier was first demonstrated at a repetition rate of 100 kHz. While this regime is attractive for high-repetition-rate and high-average-power operation, extending the investigation toward lower repetition rates is also important, as it allows addressing higher pulse energies. For CW-pumped amplifier systems, output stability becomes a critical issue due to energy bifurcation behavior. Such bifurcation occurs when the repetition rate exceeds the inverse upper-state lifetime of the gain medium, leading to pulse-to-pulse energy fluctuations in the output [119]. For $Ho^{3+}$-doped materials, this condition places the critical regime in the kHz range, which therefore requires careful verification. The bifurcation behavior of a CW-pumped Ho:CALGO amplifier has been studied using a numerical simulation in [120], where it was shown that stable, high-energy pulse extraction is possible at 1 kHz, indicating that this regime is viable for further energy scaling. Experimentally, 2.6-mJ pulse energy at 1 kHz has been demonstrated very recently, using an updated seed source and stretcher [121]. In that work, the achievable pulse energy was primarily limited by crystal damage. Further scaling toward the 10 mJ level would be feasible by reducing the peak intensity inside the crystal, for example by enlarging the beam size or increasing the stretching ratio of the seed pulse.

Nevertheless, pulse energy scaling in Ho:CALGO is ultimately constrained by its relatively strong nonlinear and thermal properties. Compared to Ho:YLF, which is widely used in Ho-based amplifier systems, CALGO exhibits a higher nonlinear refractive index $n_2$ [122,123] and thermo-optic coefficient d$n$/d$T$. Therefore, strong nonlinearity, thermal lensing, and associated beam quality degradation are expected to impose fundamental limits on further energy scaling in bulk geometries, particularly at higher average powers and pulse energies.

In the current system, a RA was utilized as the first amplification stage to compensate for the relatively small gain cross section of Ho:CALGO, enabling a sufficient number of round-trips to saturate the gain. However, with sufficiently high seed pulse energy, a multi-pass amplifier configuration becomes an attractive alternative, which can eliminate losses of the Pockels cell in the RA cavity and simplify the overall system. In addition, it potentially serves as an efficient booster amplifier for further pulse energy enhancement.

Finally, as discussed in Section 5, the average power in bulk Ho:CALGO amplifiers is currently limited by thermal effects. This motivates the exploration of alternative gain geometries that offer superior cooling efficiency. In particular, slab amplifiers are promising due to their excellent cooling efficiency and scalability, and thin-disk geometry offers an attractive route toward simultaneous power and energy scaling, which will be discussed in more detail in the following section.

*6.4 Average power scaling*

Ultrafast Ho:CALGO lasers have been demonstrated only in bulk geometry so far. For further power scaling toward hundred-watt to kilowatt levels, the thermal load in the gain material will be a key challenge, leading to cavity instability due to thermal lensing or damage to the gain crystal. Moreover, high nonlinear refractive indices $n_2$ of CALGO impose another challenge in managing nonlinearity in high-energy amplifiers. A thin-disk gain geometry offers a promising route to address both thermal and nonlinear limitations, enabling further power scaling. In the 1-μm wavelength region, the thin-disk architecture has already proven to be a transformative technology. For Yb-based lasers, it enabled a dramatic leap in average output power, extending the achievable power level by orders of magnitude [124–127]. Similar progress has begun to emerge at 2-μm wavelengths [128], where Ho:YAG thin-disk oscillators have demonstrated excellent performance [64–66], however, only a limited number of reports exist so far, and the potential of thin-disk technology in this wavelength range remains widely unexplored.

Nevertheless, implementing a Ho-based thin-disk laser presents distinct challenges compared to Yb-based systems. While the $Yb^{3+}$ ion has only two energy levels for transition, the $Ho^{3+}$ ion exhibits a more complex energy-level structure, leading to unwanted transitions such as ETU. That significantly increases heat generation, which requires careful thermal management. Ideally, higher $Ho^{3+}$ doping concentrations would compensate for the short gain length intrinsic to the thin-disk geometry, however, higher doping levels increase the probability of ETU and further increase the thermal load.

These considerations motivate above-mentioned investigations of the transitions and energy-transfer dynamics in Ho:CALGO, as well as careful optimization of doping concentration and disk dimension utilizing thermal simulations. With a properly engineered combination of material parameters, thermal design, and cavity configuration, Ho:CALGO thin-disk lasers have the potential to emerge as a next-generation, high-power ultrafast source in the 2-μm region.

In addition to geometric strategies such as the thin-disk concept, cryogenically cooled laser operation presents another promising route for further power scaling. Cryogenic cooling has shown remarkable success in both power and energy scaling across a variety of gain materials [129], because the reduced temperature increases the mean free path of phonons, thereby enhancing thermal conductivity and reducing the thermal expansion coefficient. Additionally, the absorption and emission cross sections of rare-earth ions increase at low temperatures, further improving the gain. For three-level lasers, reduced temperature suppresses reabsorption, effectively shifting toward a quasi-4-level laser system and reducing the laser threshold.

For 2-μm lasers, cryo-cooled Ho-based lasers have already demonstrated strong potential in CW [130,131], Q-switched lasers [132,133] and CPAs [68] based on Ho:YAG, Ho:YLF, Ho:YAP, and Ho:$CaF_2$. These results indicate that cryogenic operation is straightforward for mitigating thermal limitations that commonly restrict power scaling in room-temperature laser operation.

However, cryogenic cooling also has a well-known drawback: the spectral bandwidth narrows significantly as the Boltzmann population of Stark levels collapses toward the lowest sublevels. In ordered host crystals such as YAG or YLF, this narrowing severely limits the available gain bandwidth [26,134], making broadband or ultrafast operation difficult or even impossible.

In contrast, disordered materials including CALGO exhibit strong inhomogeneous broadening as shown in Fig. 8, which enables them to maintain a broad emission spectrum even at cryogenic temperatures. This is particularly attractive for ultrafast laser development, as it preserves the bandwidth necessary for sub-picosecond or femtosecond pulse generation and amplification. However, cryogenically-cooled laser operation using such disordered host materials remains widely unexplored, and only CW operation has been reported in a Tm,Ho codoped disordered garnet crystal laser so far [135]. Investigating cryogenic operation of disordered materials such as Ho:CALGO could open a new pathway toward high-power ultrafast lasers, combining the thermal advantages of low-temperature operation with the spectral benefits of inhomogeneous broadening.

*6.4 Potential for laser operation at 3 µm*

Building on the successful demonstration of CALGO as a laser host crystal for ultrashort-pulse, high-power lasers, a clear shift of interest has been observed from the 1-µm spectral range, typically addressed by $Yb^{3+}$ ions, toward the 2-µm region governed by $Tm^{3+}$,$Ho^{3+}$-codoped or singly $Ho^{3+}$-doped crystals as shown above. It is therefore interesting to extend this exploration further, toward wavelengths approaching 3 µm. This spectral region marks the transition to the mid-infrared and simultaneously represents the upper performance limit of conventional oxide laser gain media. The vibronic properties of CALGO suggest that laser action may remain feasible in the 2–3 µm range, as recently demonstrated by 2.3-µm laser operation in Tm:CALGO exploiting the $^3H_4 \rightarrow {^3H_5}$ transition [136] (even though the emitting state is prone to a significant non-radiative path via multiphonon relaxation). It is therefore natural to investigate the less commonly explored, yet highly compelling, $Ho^{3+}$ transition $^5I_6 \rightarrow {^5I_7}$, which gives rise to laser emission around 2.9 µm. Laser sources emitting around this wavelength are of interest for spectroscopy, laser surgery, material processing and environmental monitoring. One of the difficulties for developing 2.9 µm Ho lasers is the self-terminating nature of this transition due to the longer lifetime of the lower laser level causing the bottleneck effect. It can be overcome by cascade lasing, $^5I_6 \rightarrow {^5I_7}$ and $^5I_7 \rightarrow {^5I_8}$, draining the population of the intermediate metastable manifold, and via various codoping schemes. Codoping of laser crystals with $Yb^{3+}$ (donor), $Ho^{3+}$ (acceptor), and $Eu^{3+}$ or $Tb^{3+}$ (quencher) ions was suggested to access conventional pump wavelengths around 0.98 µm due to $Yb^{3+}$ absorption followed by an efficient $Yb^{3+} \rightarrow Ho^{3+}$ energy transfer, as well as to reduce the lifetime of the terminal laser level by selective de-excitation in the ensemble of lower-lying closely separated manifolds of $Eu^{3+}$ or $Tb^{3+}$ ions [137]. However, such complex codoping requires a precise balance of various phonon-assisted energy-transfer processes which would inevitably lead to additional parasitic heat deposition paths.

Building up on the concept exploited in the present paper for 2.1-µm Ho:CALGO lasers (in-band pumping), direct excitation of $Ho^{3+}$ ions to the $^5I_6$ state appears to be the most viable option. The $^5I_8 \rightarrow {^5I_6}$ transition in absorption corresponds to a peak $\sigma_{abs}$ = 1.69×10$^{-20}$ cm$^2$ at 1145 nm and an absorption bandwidth of 14 nm (for π-polarized light), a wavelength which can be addressed by Raman-shifted Yb-fiber lasers. Figure 16 depicts the SE cross-sections for the $^5I_6 \rightarrow {^5I_7}$ transition at 2.9 µm for π and σ light polarizations, calculated using the F-L formula and the radiative transition probabilities obtained using the J-O theory. The peak $\sigma_{SE}$ amounts to 5.00×10$^{-20}$ cm$^2$ at 2861 nm for π-polarization and 2.74×10$^{-20}$ cm$^2$ at 2862 nm for σ-polarized light.

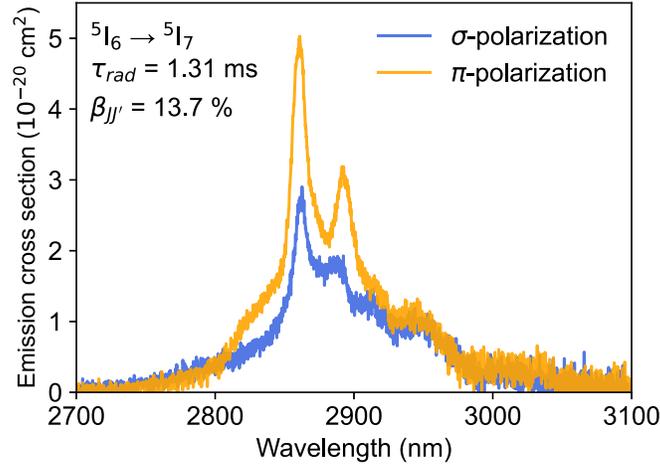

Fig. 16. Emission properties of Ho:CALGO at ~2.9 µm: room-temperature stimulated-emission, $\sigma_{SE}$, cross-sections for the $^5I_6 \rightarrow {}^5I_7$ transition of $Ho^{3+}$ ions for $\pi$ and $\sigma$ light polarizations.

## 7. Conclusion

Ho:CALGO has emerged as one of the most promising gain materials for high-power ultrafast lasers in the 2-µm wavelength range. The unique combination of a broad and flat gain profile and superior thermo-mechanical properties enables outstanding laser performance in mode-locked lasers and CPA systems. These laser systems demonstrate exceptionally short pulse durations among all Ho-based laser systems at high average power levels. These achievements clearly establish Ho:CALGO as a leading platform for next-generation mid-infrared ultrafast lasers.

Despite this rapid progress, there is further room for improvement. Advanced spectroscopic investigations, including detailed studies of energy-transfer processes, will be essential for a deep understanding of Ho:CALGO and enabling modeling of laser operation. Such models will facilitate the optimization of doping concentration, crystal dimensions, geometry, cavity design, and so on. These improvements will increase laser efficiency and reduce thermal load, and also expand the accessible operating regimes for future high-power systems toward the hundred watts level. Further work on pulse shortening is possible aiming for few-cycle pulse durations as well.

From a technological point of view, advanced gain geometries and thermal-management strategies offer the opportunity to scale Ho:CALGO lasers to even higher powers and pulse energies. Thin-disk geometry is promising to provide a breakthrough in 2-µm laser systems once the specific challenges of Ho-doped materials, such as ETU and its additional heat generation are systematically analyzed and properly managed. In parallel, slab geometry also offers a complementary scaling approach, benefiting from efficient heat extraction. Furthermore, cryogenic laser operation will enable the extension for power and energy scaling, benefiting from improved thermos-mechanical and thermo-optic properties, reduced reabsorption, and enhanced cross sections at low temperatures. The disordered nature of CALGO is particularly advantageous as it can maintain a broad bandwidth due to inhomogeneous line broadening, thus preserving ultrafast capability even under cryogenic cooling.

In summary, Ho:CALGO uniquely combines broad gain bandwidth, robust thermal properties, and compatibility with both advanced geometries and cryogenic operation. With spectroscopic analysis, development of laser models, and the integration of laser-engineering technologies, Ho:CALGO-based systems are well positioned to shape the next generation of high-power ultrafast sources in the 2-μm region.

**Funding.** European Research Council (ERC) under the European Union's Horizon 2020 research and innovation programme (grant agreement No. 805202 - Project Teraqua), European Research Council (ERC) under the European Union's HORIZON-ERC-POC programme (Project 101138967 - Giga2u), Deutsche Forschungsgemeinschaft (DFG, German Research Foundation) under Germanys Excellence Strategy – EXC-2033 – Projektnummer 390677874 - RESOLV. French Agence Nationale de la Recherche (ANR-22-CE08-0031, FLAMIR). Région Normandie, France (Contrat de plan État-Région, CPER)

**Acknowledgment.** The authors gratefully acknowledge Dr. Christoph Liebald and Dr. Daniel Rytz from Coherent GmbH for fruitful discussions. We also thank Rayven Laser GmbH, the spin-off company dedicated to the commercialization of Ho:CALGO-based laser technology, for their support and collaboration. These results are part of a project that has received funding from the European Research Council (ERC) under the European Union's Horizon 2020 research and innovation programme (grant agreement No. 805202 - Project Teraqua). These results are part of a project that has received funding from the European Research Council (ERC) under the European Union's HORIZON-ERC-POC programme (Project 101138967 - Giga2u). Funded by the Deutsche Forschungsgemeinschaft (DFG, German Research Foundation) under Germanys Excellence Strategy – EXC-2033 – Projektnummer 390677874 - RESOLV. This work was also funded by French Agence Nationale de la Recherche (ANR-22-CE08-0031, FLAMIR) and Région Normandie, France (Contrat de plan État-Région, CPER).

**Disclosures.** The authors declare no conflicts of interest.

**Data availability.** Data underlying the results presented in this paper can be obtained from the authors upon reasonable request.